%% file: Main.tex
\documentclass[sigconf]{acmart}
\usepackage[linesnumbered,ruled,vlined]{algorithm2e}

\SetCommentSty{mycommfont} %

\AtBeginDocument{%
  }

\copyrightyear{2025}
\acmYear{2025}
\setcopyright{rightsretained}
\acmConference[CHIWORK '25]{CHIWORK '25: Proceedings of the 4th Annual Symposium on Human-Computer Interaction for Work}{June 23--25, 2025}{Amsterdam, Netherlands}
\acmBooktitle{CHIWORK '25: Proceedings of the 4th Annual Symposium on Human-Computer Interaction for Work (CHIWORK '25), June 23--25, 2025, Amsterdam, Netherlands}
\acmPrice{}
\acmDOI{10.1145/3729176.3729197}
\acmISBN{979-8-4007-1384-2/25/06}

\usepackage{appendix}
\usepackage{fancybox}
\usepackage{listings}
\usepackage{subcaption}
\usepackage{soul}
\definecolor{DarkGoldenrod}{rgb}{0.72, 0.53, 0.04}

\newcommand{\toolname}{\textsc{tAIfa}}

\newcommand{\promptbox}[1]{
  \begin{center} 
    \doublebox{%
      \begin{minipage}{.95\columnwidth} 
        \vspace{5pt} 
        #1
        \vspace{5pt} 
      \end{minipage}%
    }
  \end{center}
}

\begin{document}

\title[tAIfa]{tAIfa: Enhancing Team Effectiveness and Cohesion with AI-Generated Automated Feedback}

\author{Mohammed Almutairi}
\email{malmutai@nd.edu}
\affiliation{%
  \institution{University of Notre Dame}
  \city{Notre Dame}
  \state{Indiana}
  \country{USA}
  \postcode{46556}
}

\author{Charles Chiang}
\email{cchiang3@nd.edu}
\affiliation{%
  \institution{University of Notre Dame}
  \city{Notre Dame}
  \state{Indiana}
  \country{USA}
  \postcode{46556}
}

\author{Yuxin Bai}
\email{yuxin.bai@stu.xjtu.edu.cn}
\affiliation{%
  \institution{University of Notre Dame}
  \city{Notre Dame}
  \state{Indiana}
  \country{USA}
  \postcode{46556}
}

\author{Diego Gomez-Zara}
\email{dgomezara@nd.edu}
\affiliation{%
  \institution{University of Notre Dame}
  \city{Notre Dame}
  \state{Indiana}
  \country{USA}
  \postcode{46556}
}
\renewcommand{\shortauthors}{Almutairi et al.}

\begin{abstract}
\input{00_Abstract}
\end{abstract}





\begin{teaserfigure}
    \centering
    \includegraphics[width=0.9\textwidth]{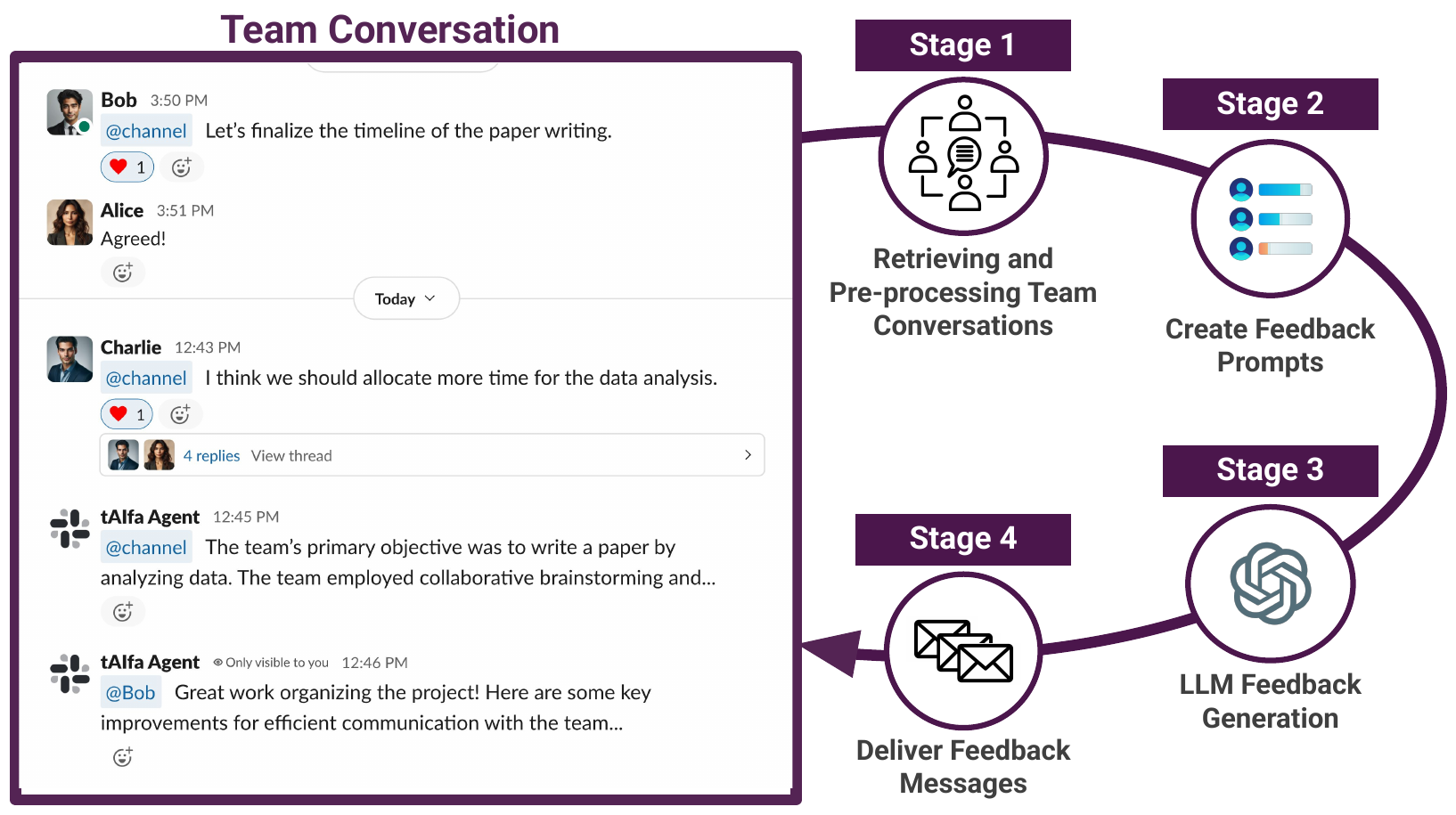}
    \caption{Overview of the \toolname{} agent implemented in Slack. Stage \textbf{\textcircled{1}}: The system retrieves and structures team conversations. Stage \textbf{\textcircled{2}}: It evaluates team dynamics with rule-based metrics and LLM-based analysis. Stage \textbf{\textcircled{3}}: The system employs an LLM to generate feedback messages for individual and team messages. Stage \textbf{\textcircled{4}}: The system delivers individual feedback messages as private messages (``Only visible to you'') and group feedback messages tagging all the channel's members.}
    \label{fig:High-level system architecture}
\end{teaserfigure}

\maketitle

\input{01_Introduction}
\input{02_RelatedWork}

\input{03_FormativeStudy}

\input{04_System}
\input{06_UserStudy}

\input{07_Results}

\input{08_Discussion}
\input{09_Conclusion}
\input{10_EthicsStatement}

\begin{acks}
This project was supported by Slack's 2024 Workforce Lab Academic Grant Program and Microsoft Research's Accelerating Foundation Models Research Program.
\end{acks}

\bibliographystyle{ACM-Reference-Format}
\bibliography{0_Citation}

\appendix
\input{Appendix/FormativeStudyPrompts}
\input{Appendix/LSM}
\input{Appendix/Task}
\input{Appendix/UserStudy}



\end{document}

%% file: 00_Abstract.tex
Providing timely and actionable feedback is crucial for effective collaboration, learning, and coordination within teams. However, many teams face challenges in receiving feedback that aligns with their goals and promotes cohesion. We introduce \toolname{} (``Team AI Feedback Assistant''), an AI agent that uses Large Language Models (LLMs) to provide personalized, automated feedback to teams and their members. \toolname{} analyzes team interactions, identifies strengths and areas for improvement, and delivers targeted feedback based on communication patterns. We conducted a between-subjects study with 18 teams testing whether using \toolname{} impacted their teamwork. Our findings show that \toolname{} improved communication and contributions within the teams. This paper contributes to the Human-AI Interaction literature by presenting a computational framework that integrates LLMs to provide automated feedback, introducing \toolname{} as a tool to enhance team engagement and cohesion, and providing insights into future AI applications to support team collaboration.

%% file: 01_Introduction.tex
\section{Introduction}
Providing feedback to team members is essential for enabling effective collaboration, promoting learning, enhancing coordination, and addressing weaknesses \cite{handke2022unpacking, penarroja2015team}. Despite its importance, one main challenge for many teams is receiving constructive, opportune, and actionable feedback \cite{VanThielen2018}. While managers, mentors, and leaders often serve as the primary sources of feedback and guidance, their effectiveness depends on their availability, interpersonal skills, and the trust levels within their teams \cite{hsieh2018exploring,henderson2005developing}. Fostering feedback is even more challenging in remote and hybrid teams since managers' and leaders' interactions with employees are less visible to the entire team compared to co-located teams \cite{marlow2017communication,hoch2014leading}. Remote teams must constantly synchronize not just tasks but also cognitive and social states to maintain their performance and cohesion \cite{deSouza2022,Hinds2003}. 

To address this issue, computer science researchers have developed automated tools that provide feedback to teams, aiming to address ineffective group dynamics (e.g., low participation or collaboration) and guide members toward ideal norms of collaboration \cite{leshed2009visualizing, tausczik2013improving, mcduff2016affdex, boyd2016saywat, faucett2017should}. For example, He et al. \cite{he2017two} designed a feedback tool to enhance cross-cultural team empathy using linguistic analysis and sentiment detection. Similarly, Sharon et al. \cite{zhou2018search} developed a chatbot called `DreamTeam' that optimizes team compositions dynamically by adjusting norms, hierarchies, and communication styles employing a multi-armed bandit framework. Lastly, Cao et al. \cite{cao2021my} studied team viability by analyzing interactions through existing machine learning models to facilitate early interventions.

Although many feedback tools and frameworks exist, most solutions depend on pre-coded recommendation messages, basic text analytics (e.g., sentiment, keyword usage), or limited training datasets that may not generalize effectively across different domains and contexts \cite{rivers2017data, hadi2023evaluating}. Building on prior research in collaborative team systems and automated feedback, this study aims to leverage state-of-the-art AI models to provide automated, timely feedback. Recent advancements in Large Language Models (LLMs) provide new opportunities to develop feedback systems that are more contextually aware, personalized to individual and team needs, and adaptable to a broader range of tasks and diverse team environments \cite{zhang2024personalization}.

In this study, we present \toolname{} (``Team AI Feedback Assistant''), an innovative LLM-based agent designed to deliver tailored feedback to team members working online. This system analyzes team interactions to identify areas for improvement and generates personalized feedback for individual members and the team as a whole. Our primary contribution is a computational framework that employs LLMs to analyze team conversations without using preexisting user data to generate feedback messages that enhance team effectiveness and engagement. By integrating LLM reasoning capabilities with empirical findings from small group research, \toolname{} leverages both linguistic and behavioral indicators to generate personal feedback messages that promote communication and cohesion. Furthermore, this work extends existing automated feedback systems by addressing cognitive and social dimensions of team collaboration while providing highly customizable guidance to enhance team effectiveness. 

To evaluate this proposed LLM-based framework, we conducted a between-subjects study with 54 participants, who assembled 18 teams to complete three problem-solving tasks in a Slack workspace. We randomized teams into two conditions: teams working without receiving any feedback (i.e., control) and teams receiving feedback from \toolname{} by the end of each task (i.e., treatment). Across three rounds, we analyzed teams' transcripts to evaluate their group dynamics and surveyed participants about their team experiences and collaborative work. Our results indicate that teams receiving AI-generated feedback engaged in longer conversations and exchanged turns more frequently than teams without receiving feedback, suggesting more dynamic and balanced interactions. 

This paper makes the following contributions: (1) a formative study investigating how users respond to different feedback messages generated by an LLM, (2) a novel LLM-based agent that enhances team participation through individual and group feedback, and (3) a lab study with 54 participants that demonstrates \toolname{} positive influence on team communication patterns. The source code of \toolname{} is available at \url{https://github.com/RINGZ-Lab/taifa}.

%% file: 02_RelatedWork.tex
\section{Related Work} 
\label{RelatedWork}

\subsection{Feedback in Teams}
Feedback is the act of providing team members with information about how well they are performing a task and suggesting ways to improve \cite{hackman1971employee}. Receiving feedback helps members enhance their performance and cohesion by changing their actions, motivations, and behaviors \cite{masthoff2024towards,gomez2020taxonomy}. Feedback can include subjective or objective information, incorporating aspects such as conversational insights or performance evaluations. Moreover, it can be provided at the individual (e.g., personalized instructions to a person) and group level (e.g., guidance for the whole team) \cite{nadler1979effects}. Since tasks are rarely static and often unfold in different cycles \cite{deshon2004multiple}, leaders can monitor team members' performance and diagnose problems to provide timely feedback, enhancing team learning and development \cite{kozlowski2008developing}. 

Previous research categorizes team feedback into two main types: \textit{recipient-focused} and \textit{time-focused} feedback \cite{deeva2021review}. The first type of feedback focuses on the information and guidance the recipient should receive, ranging from messages for each team member to messages for the entire team as a whole. In contrast, the second type of feedback focuses on the appropriate information a team should receive, given the moment of the task, whether before, during, or after task completion. 

Receiving feedback during the task has several benefits. First, it can help team members focus on understanding the requirements and execution of the tasks, directing or putting their efforts back on track toward the team's goal \cite{nikiforow2021contextual}. Second, it can provide members with a sense of accomplishment as they can address and manage the feedback received immediately \cite{nikiforow2021contextual}. However, receiving feedback too often can increase distractions and shift members' attention from focusing on task-specific objectives to monitoring performance metrics instead  \cite{butler2007effect}. It can also fail to include contextual factors of the task, which could be relevant or necessary later \cite{walsh2009concurrent}.

In this work, we examine how a computational system can deliver timely feedback to teams by leveraging the information members receive and the required immediacy to enhance their work.

\subsection{Real-time Feedback Systems}
\label{real-time-generated-systems}
Artificial intelligence (AI) has gained significant interest in automating team tasks and interactions among team members \cite{parker2022automation}. In particular, AI-generated feedback can help organizations automate performance evaluation, streamline team monitoring, and provide continuous improvement suggestions without the need for constant human intervention \cite{daryanto2025conversate, benharrak2024writer, munyaka2023decision}. Moreover, with the increasing number of teams in organizations and institutions, automating feedback can become an important tool to address scalability issues. 

Previous studies have examined collaboration tools that provide feedback to teams while they complete their tasks \cite{keuning2018systematic}. For example, Faucett et al. \cite{faucett2017should} introduced `ReflectLive,' a system that offers real-time feedback on non-verbal interactions of clinicians during video consultations. This system monitors behaviors---such as eye gaze and interruptions---to provide real-time feedback on a dashboard, enabling clinicians to be more aware of their behavior and interactions with patients. Boyd et al. \cite{boyd2016saywat} developed `SayWAT,' a real-time feedback tool on prosody during face-to-face interactions. Implemented on Google Glass, this tool provides visual feedback on atypical speech patterns, such as pitch and volume irregularities. Similarly, Samrose et al. \cite{samrose2020immediate} developed a video chat platform that gives real-time feedback on interruptions, volume, and facial emotion during team discussions. Users receive private insights into their talk duration, emotional labels, and interruption frequencies, promoting frequent behavior adjustments. 

While these existing tools offer valuable assistance, many rely on simple heuristics and overlook the deeper context and semantics of team interactions. This simplicity can lead users to ``game'' the system, making superficial adjustments to their behavior without achieving meaningful improvement \cite{kyrilov2015binary}. Moreover, the interpretation of these rules-based team dynamic values can vary depending on the task, limiting their generalizability \cite{faucett2017should, leshed2009visualizing}. With \toolname{}, our goal is to provide AI-generated feedback that encourages teams to reflect on and refine their communication patterns over time. 

\subsection{Large Language Models in Team Dynamics}
\label{llm-in-teams}
Several studies have explored how Large Language Models (LLMs), such as OpenAI's GPT \cite{achiam2023gpt} and Meta's LLaMA \cite{touvron2023llama}, can provide assistance and guidance to teams. Some frequent applications include idea generation \cite{he2024ai}, decision-making \cite{ataguba2025exploring, wang2021deep}, and collaborative problem-solving \cite{suh2021ai, song2022decoding, ashktorab2021effects}.

A promising application for LLMs is automating team feedback. Katz et al. \cite{katz2023exploring} found that ChatGPT could identify relevant feedback themes from student feedback comments with high precision. Similarly, Wei et al. \cite{wei2024improving} explored LLM-powered virtual assistants that provided learners with real-time feedback, enabling higher participation levels, improved learning rates, and increased emotional satisfaction among students. In another example, Chiang et al. \cite{chiang2024enhancing} used an LLM-powered devil's advocate to challenge AI-generated recommendations provided to teams, showing that teams made more accurate decisions when the system dynamically opposed the AI-generated suggestions and responded to the ongoing conversations. Lastly, Sidji et al. \cite{sidji2024human} investigated the impact of an LLM-powered AI assistant on team dynamics in cooperative games. Although the assistant fostered creative thinking by providing a creative starting point for clue generation, many players became more focused on understanding the LLM's reasoning than on aligning with their teammates' thought processes. 

Unlike previous systems, we build \toolname{} grounded in team dynamics research to deliver automated and timely feedback to individuals and teams, enabling them to reflect on their progress and interactions as they collaborate.

%% file: 03_FormativeStudy.tex
\section{Feedback Messages Design}
\label{prompt-design}
In this section, we detail our approach to designing team feedback messages. First, we outline how we selected and operationalized relevant team communication metrics. Next, we describe how we designed and refined LLM prompts to generate actionable feedback. Finally, we present a formative evaluation to assess the clarity, usefulness, and effectiveness of these prompts. 

\subsection{Communication Metrics Development}
To develop \toolname{}, we focus on providing feedback based on team communication patterns that can trigger reflection and behavioral changes. A key design goal is to generate feedback exclusively from ongoing conversations, avoiding relying on preexisting user data. By drawing primarily on the immediate context, we aim to create a tool that is generalizable across diverse domains and minimizes users' preexisting data. 

We prompt an LLM to judge the team dynamics and interactions based on the teams' transcripts. Drawing on small-group studies, we employ seven team communication metrics to guide the LLM's evaluation (Table \ref{tab:team_dynamics_evaluation}). These metrics evaluate participation (i.e., the extent to which participants contribute and exchange information) and cohesion (i.e., the extent to which team members maintain unity to achieve their goals) \cite{downing1958cohesiveness, tan2022communication}.

\input{tables/team_metrics}

\subsubsection{Text Analytics Metrics.}
Our first set of metrics involved traditional text-based metrics that quantify linguistic patterns in the transcripts. These have specific scales to assess team dynamics.

\paragraph{Language Style Matching (LSM)} It measures how similarly two or more team members use ``function words'' (e.g., pronouns, articles, conjunctions, and auxiliary verbs) when they communicate \cite{aafjes2020language}. Previous research shows that high LSM levels reflect conversational partners aligning their language styles to achieve a common goal \cite{aafjes2020reciprocal, ireland2010language}. \toolname{} analyzes the frequency of function words used by each team member and then compares the word distributions across messages to determine similarity scores between participants.

\paragraph{Sentiment.} The emotional tone of a conversation influences how messages are interpreted by team members. Since emotion naturally emerges during discussions, sentiment analysis helps identify emotional contagion and its impact on cohesion \cite{kane2023emotional}. \toolname{} measures the sentiment of the text, categorizing words into predefined positive, negative, or neutral groups and assigning a score to each word. The overall sentiment score is then calculated by averaging these individual scores across the conversation. This final score shows if the group discussion was positive, negative, or neutral.
        
\paragraph{Team Engagement.} It evaluates the level of contribution of team members to discussions and decision-making \cite{yoerger2015participate}. High-engaged teams show more balanced participation, with all members actively sharing diverse perspectives, than low-engaged teams, which have only a few members participating in the discussions and decisions. \toolname{} measures engagement by calculating the number of words contributed by each member relative to the overall discussion \cite{tausczik2010psychological}.

\subsubsection{Contextual Metrics.}
This second set of metrics centered on contextual, latent constructs from the teams' transcripts. Rather than relying on specific text-based formulas, we prompted an LLM to assess them by interpreting the content of the conversation, thus providing an in-depth, contextual analysis of the team members' interactions. Inspired by LLM-as-a-judge frameworks and strong performance in linguistic classification and semantic labeling tasks \cite{latif2024systematic, pan2024human, gu2024survey}, we prompted an LLM to utilize these capabilities to evaluate the transcripts using the following four metrics.

\paragraph{Transactive Memory System (TMS)} TMS represents the way teams coordinate by leveraging each member's knowledge through the development of a shared cognitive system \cite{liang1995group}. Teams with high TMS scores anticipated better others' roles and expertise---leading to higher coordination and less frequent communication---than those with low TMS scores \cite{marques2013and}. \toolname{} examines this factor to determine whether team members recognize each other's expertise in the decision-making process to enhance teamwork, as successful teams effectively utilize each member's expertise \cite{garrett2009six}. \toolname{} uses prompts to instruct an LLM to analyze a team's conversations and evaluate how often team members correctly identified and referred to the appropriate teammate for specific tasks or expertise.

\paragraph{Collective Pronouns.} Counting the number of collective pronouns (e.g., ``we'', ``our'') is a well-established linguistic marker of shared team identity \cite{gonzales2010language}. A higher frequency of these pronouns often indicates a stronger sense of collaboration within teams. However, using simple keyword matching can miss the contextual nuances of group messages \cite{shin2021constrained}. For example, the sentence \textit{``You and I make a great team''} lacks explicit collective pronouns but still conveys a collaborative sentiment. To address this limitation, \toolname{} instructs the LLM to interpret how frequently team members framed their contribution as part of a collective effort \cite{chen2025semantic}.
    
\paragraph{Communication Flow.} This metric assesses how team members interact, identifying interruptions, instances of ignoring others, and delayed responses \cite{klunder2016communication}. Frequent interruptions might indicate poor coordination while ignoring others, or delayed responses can indicate disengagement within the team. Traditional approaches employ fixed temporal thresholds to define interruption or lapses in interaction. However, these methods often fail to account for the nuance of the conversation \cite{chen2025semantic}. To address this, \toolname{} prompts the LLM to assess the flow of communication by jointly analyzing both timestamps and semantic content. This enables \toolname{} to distinguish between a meaningful pause (e.g., a natural transition between two individuals) and actual interruptions in the conversation \cite{more2024analyzing}.

\begin{figure*}[!htb]
    \centering
    \includegraphics[width=\linewidth]{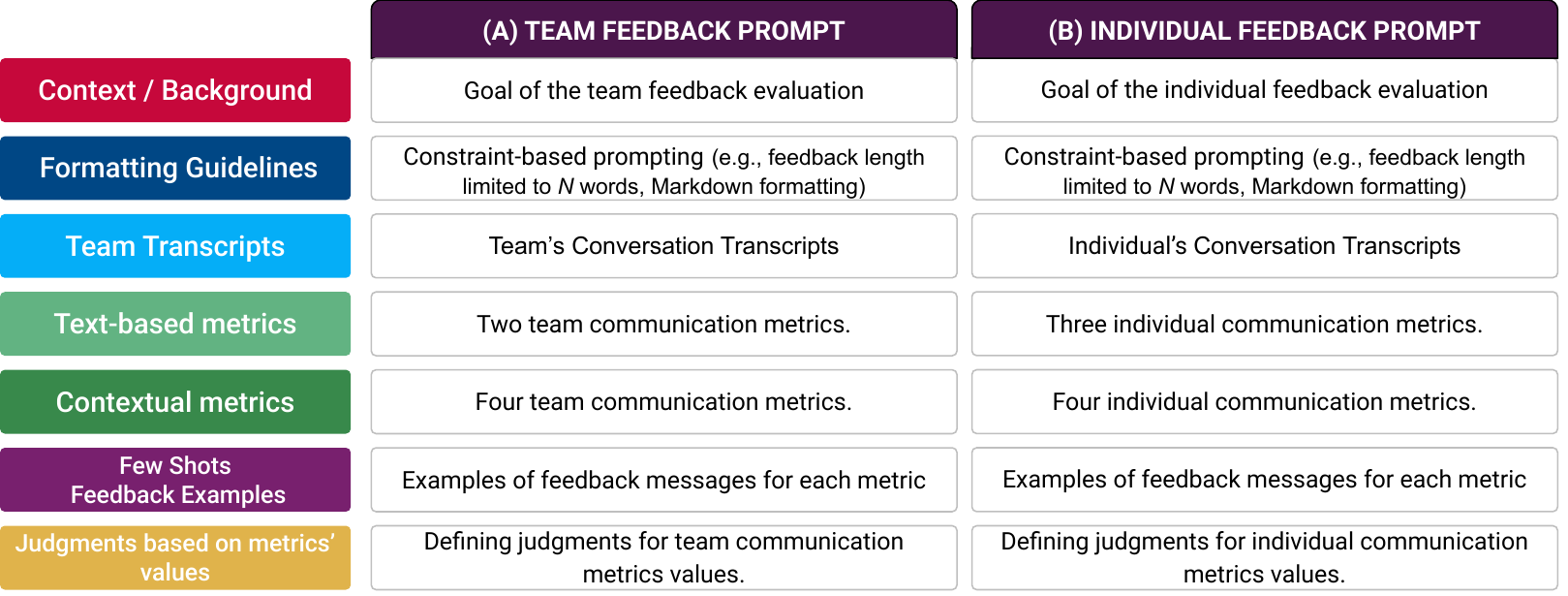}
    \caption{Anatomy of Team and Individual AI Feedback Prompts. A structured breakdown of prompts used to generate feedback for (A) teams and (B) individuals. The full team and individual prompts can be found in Appendix \ref{apandxPrompts}.}
    \label{fig:prompt-layers}
\end{figure*}

\paragraph{Topic Coherence.} Our final metric evaluates the alignment between team members' conversations and the team's objectives. Traditional methods, such as topic modeling, can extract the general themes from a conversation, but often struggle to capture nuanced, context-dependent topics, as well as topic shifts within conversations \cite{gupta2024comprehensive, gillings2023interpretation}. To overcome this limitation, \toolname{} prompts the LLM to compare the team's task description with the team transcript, enabling it to assess the semantic coherence between the ongoing conversation and the team's defined goal \cite{mu2024large}.

\subsection{Prompt Design}
Building on the identified communication metrics, we piloted nine distinct prompts to examine how varying the amount of contextual information provided to the LLM affected the quality and effectiveness of the automated feedback messages \cite{bandi2025evaluation, onan2025assessing}. To systematically explore prompt design strategies, we varied the prompts along two dimensions: the length of the feedback message shown to users (i.e., short, medium, long) and the level of contextual information derived from the metrics provided to the LLM (i.e., low, medium, high). Regarding the message length, we created prompts limiting the number of words and details. \textit{Short feedback} was limited to approximately 100 words and mostly focused on high-level guidance. \textit{Medium feedback} was limited to approximately 200 words and offered additional contextual examples and elaboration. Lastly, \textit{long feedback} messages were up to approximately 300 words and provided in-depth explanations of team performance and areas for improvement with examples drawn from the conversation.

For the second dimension, we designed prompts with varying levels of contextual information by incrementally adding information blocks to the prompts (Figure \ref{fig:prompt-layers}). In the \textit{low context} condition, prompts included only the teams' conversation transcripts without any defined team communication metrics or input data. This setup relied on the LLM's reasoning capabilities to infer communication patterns, analyze interactions, and generate feedback with high autonomy. In the \textit{medium context} condition, we augmented the low-context prompt with the communication metrics and included few-shot examples to guide the feedback generation process. Lastly, in the \textit{high context} condition, we extended the prompt by incorporating explicit judgments based on the metrics' values. This approach constrained the LLM's interpretive capability, directing it to generate messages aligned with the predefined judgments. 

\input{tables/prompts}

\subsection{Formative Study}
\label{Formative Study}
We conducted a formative study to assess the usefulness of the feedback generated by each prompt. We recruited 30 participants on Prolific and compensated them with \$8 USD for completing the study, which took approximately 40 minutes. The payment was set according to the platform's recommended rate of \$12 USD per hour. Fourteen participants self-identified as male and 16 as female. All participants reported being fluent in English. The participants' ages range from 20 to 62 years, with a mean age of 36.6 years. The study was reviewed and approved by our institution's Institutional Review Board ({\#24-02-8358}).

\subsubsection{Procedure.}
Prolific participants were directed to a survey developed by the research team. On the first page, participants read a 5-minute simulated conversation in which a team worked to resolve a \textit{survival task}, a fictitious scenario where the team must decide which resources to prioritize to survive in an isolated environment \cite{braley2018gap}. Participants were instructed to focus on one team member's messages while performing the task. The website presented the messages on the screen according to their original timestamps--- emulating a real-time conversation---and showed the team's scores once the simulated conversation finished. 

Based on the prompts described in Table \ref{tab:llm_dependency}, the system displayed nine LLM-generated feedback messages for the selected member and another nine feedback messages directed to the team. The messages were displayed in random order to control for order effects, and the survey displayed one feedback message per page. Participants assessed the quality of these messages using the following eight evaluation criteria on a five-point Likert scale: (1) clarity and understanding, (2) feedback message satisfaction, (3) level of detail, (4) comprehensiveness, (5) accuracy and reliability, (6) actionable feedback, (7) task relevance, and (8) engagement. They could also offer open-ended comments to provide more feedback and observations (more details of the survey in Appendix \ref{appendix: Formative Study Prompts}).

\subsubsection{Results.}
\label{formative_study_section}
We first analyzed participants' responses to identify inconsistent scoring using a standard deviation threshold ($\sigma < 0.1$). Four participants were excluded due to inconsistent responses, resulting in a final sample of 26 participants for analysis. Based on these responses, we calculated the reliability of these items using Cronbach's alpha across all the prompts. The items showed high consistency, ranging from 0.84 to 0.94. Therefore, we averaged them to calculate a single score per prompt.

\begin{figure}
    \centering
    \includegraphics[width=\linewidth]{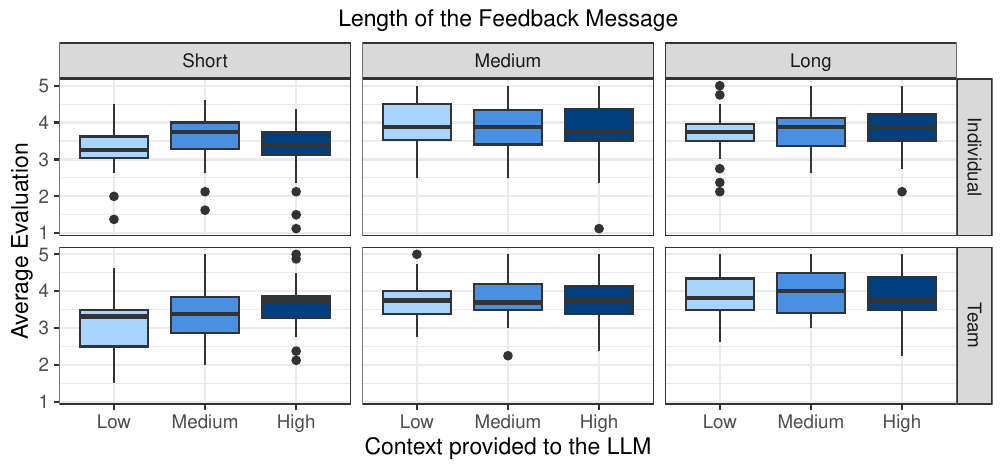}
    \caption{Average ratings of the prompts for the individual and team feedback messages.}
    \Description{Average ratings of the prompts for the individual and team feedback messages}
    \label{fig:FormativeStudyResult}
\end{figure}

Figure \ref{fig:FormativeStudyResult} shows the average ratings of the prompts for both individual and team feedback messages, categorized by the length and context-level dimensions. Overall, the length of the messages significantly influenced the ratings of individual feedback messages (Two-way ANOVA, $F(3,230)=12.30, p<0.001$) as well as for team feedback messages ($F(1,230)=27.48, p<0.001$). Moreover, incorporating more information in the prompts also significantly influenced the ratings of the team feedback messages ($F(1,230)=4.04,p<0.05$). Among the individual feedback prompts, the medium-length message with medium context received the highest rating ($M=3.94, SD=0.68$), whereas the short-length messages received the lowest ratings. For the team feedback, the highest-rated message was long in length and medium context level in the prompt ($M=3.97, SD=0.66$), and the short-length messages were rated the lowest. 

Overall, the established length and provided context to the LLM significantly influenced how participants perceived the feedback messages. These results are consistent with previous research showing that users interacting with AI prefer detailed messages for decision-making and learning \cite{salimzadeh2024dealing,zhang2022you,ashktorab2021effects}. The high ratings for longer feedback messages demonstrate that participants valued more detailed and elaborate instructions. Moreover, incorporating team communication metrics into the prompts was beneficial since participants scored those messages higher than those without any provided metrics. Based on these results, we selected the medium-context prompts to generate medium-length individual feedback messages and long team feedback messages.

%% file: tables/team_metrics.tex
\begin{table*}[!htb]
    \centering
    \renewcommand{\arraystretch}{1.2} 
    \setlength{\tabcolsep}{5pt} 
    \small 
    \resizebox{\textwidth}{!}{ 
    \begin{tabular}{p{3cm} p{4cm} p{6cm} p{3.5cm}} 
        \hline
        \textbf{Metric} & \textbf{Input Data} & \textbf{Computational Method} & \textbf{Output Data} \\ \hline
        \multicolumn{4}{l}{\textit{\textbf{Text-Analytic Metrics}}} \\ \hline
        Language Style Matching (LSM) & Individual and team transcripts & Tokenize using POS tagging to extract function words; compare percentages across categories & LSM score (0 to 1) indicating linguistic alignment \\ \hline
        Sentiment & Individual and team transcripts & Apply VADER sentiment analysis to extract sentiment-laden emoticons & Compound sentiment score (range: -1 to 1) \\ \hline
        Team Engagement & Individual and team transcripts, total team members & Tokenize transcripts and calculate each individual's word count ratio relative to the total team word count & Percentage contribution per individual \\ \hline
        \multicolumn{4}{l}{\textit{\textbf{Contextual Metrics}}} \\ \hline
        Transactive Memory Systems (TMS) & Team transcript, team members' roles & Analyze references to expertise and delegation of subtasks & Summary of effective knowledge utilization \\ \hline
        Collective Pronoun Usage & Team transcript (and individual names for personalized feedback) & Analyze the frequency and context of collective pronouns via LLM prompts & Summary of inclusive language usage \\ \hline
        Communication Flow & Team transcript with timestamps and member metadata & Evaluate turn-taking, response delays, and interruptions& Summary of communication dynamics \\ \hline
        Topic Coherence & Team task goal, team transcript & Compare the task goal with the conversation transcript & Summary of discussion alignment with the goal \\ \hline
        
    \end{tabular}
    }
    \caption{Summary of Team Communication Metrics.}
    \label{tab:team_dynamics_evaluation}
\end{table*}

%% file: tables/prompts.tex
\begin{table*}[!htb]
    \small
    \centering
    \renewcommand{\arraystretch}{1.2}
    \begin{tabular}{p{1.25cm}p{5cm}p{5cm}p{5cm}}
        \hline
        & \textbf{Low Context} & \textbf{Medium Context} & \textbf{High Context} \\ \hline
        \textbf{Short Feedback}   
        & \textbf{Prompt ID 1} \newline  
        - $\approx$ 100 words  \newline  
        - High-level feedback  \newline  
        - Minimal structured input  \newline  
        - No team dynamic evaluation data  \newline  
        - Team discussion transcript included.  \newline  
        - Generates 2-3 examples of strengths and areas for improvement  \newline  
        & \textbf{Prompt ID 2} \newline  
        - $\approx$ 100 words  \newline  
        - High-level feedback  \newline  
        - Includes few-shot examples to infer feedback structure  \newline  
        - Team dynamics evaluation and team discussion transcript included  \newline  
        - Generates 2-3 examples of strengths and areas for improvement.  
        & \textbf{Prompt ID 3} \newline  
        - $\approx$ 100 words  \newline  
        - High-level feedback  \newline  
        - Details structured guidance to generate the feedback \newline  
        - Guided by team dynamics evaluation and team discussion transcript  \newline  
        - Generates 2-3 examples of strengths and areas for improvement. \\ \hline

        \textbf{Medium Feedback}  
        & \textbf{Prompt ID 4} \newline  
        - $\approx$ 200 words  \newline  
        - Detailed feedback with more elaboration  \newline  
        - Minimal structured input  \newline  
        - Generates 4-5 examples of strengths and areas for improvement  \newline  
        - No team dynamic evaluation data  \newline  
        - Team discussion transcript included.  \newline  
        & \textbf{Prompt ID 5} \newline  
        - $\approx$ 200 words  \newline  
        - Detailed feedback with more elaboration  \newline  
        - Includes few-shot examples  \newline  
        - Team dynamics evaluation and team discussion transcript included  \newline  
        - Generates 4-5 examples of strengths and areas for improvement.  
        & \textbf{Prompt ID 6} \newline  
        - $\approx$ 200 words  \newline  
        - Detailed feedback with more elaboration  \newline  
        - Guided by team dynamics evaluation and team discussion transcript  \newline  
        - Generates 4-5 examples of strengths and areas for improvement. \\ \hline

        \textbf{Long Feedback}  
        & \textbf{Prompt ID 7} \newline  
        - $\approx$ 300 words  \newline  
        - Detailed feedback with expanded elaboration and examples drawn from the conversation context  \newline  
        - In-depth explanations  \newline  
        - Generates 5-6 examples of strengths and areas for improvement  \newline  
        - No team dynamics evaluation data  \newline  
        - Team discussion transcript included.  \newline  
        & \textbf{Prompt ID 8} \newline  
        - $\approx$ 300 words  \newline  
        - Detailed feedback with expanded elaboration and examples drawn from the conversation context  \newline  
        - Includes few-shot examples  \newline  
        - Team dynamics evaluation and team discussion transcript included  \newline  
        - Generates 5-6 examples of strengths and areas for improvement.  
        & \textbf{Prompt ID 9} \newline  
        - $\approx$ 300 words  \newline  
        - Detailed feedback with expanded elaboration and examples drawn from the conversation context  \newline  
        - Guided by team dynamics evaluation and team discussion transcript  \newline  
        - Generates 5-6 examples of strengths and areas for improvement. \\ \hline
    \end{tabular}
    \caption{Piloted feedback message prompts based on their length and context levels. The length dimension sets the maximum number of words allowed in the message, while the context dimension sets the predefined information included in the prompt.}
    \label{tab:llm_dependency}
\end{table*}

%% file: 04_System.tex
\section{System}
We developed \toolname{} (``Team AI Feedback Assistant''), an LLM agent designed to provide immediate AI-generated feedback messages to individuals and teams. The primary goal of the \toolname{} agent is to enhance team effectiveness, cohesion, and viability for teams working online. Figure~\ref{fig:High-level system architecture} illustrates the architecture of the \toolname{} agent, detailing its four main processing stages. In Stage \textbf{\textcircled{1}}, the agent retrieves team conversations using the online communication platform's API, extracts relevant messages and team performance metrics, and pre-processes them for analysis. Stage \textbf{\textcircled{2}} involves evaluating team dynamics using the conversation metrics described in Section \ref{prompt-design}. In Stage \textbf{\textcircled{3}}, the \toolname{} agent generates AI-feedback messages using an LLM model. Finally, in Stage \textbf{\textcircled{4}}, \toolname{} delivers the AI feedback messages as private messages to individuals and public messages to the team.

In this section, we provide a detailed breakdown of each stage of the system, including how it processes the team conversations, analyzes the team dynamics, and generates and delivers the feedback.

\subsection{Stage\textbf{\textcircled{1}}: Retrieving and Pre-Processing Team Conversations}
\toolname{} retrieves all messages exchanged within group channels and participants' usernames. The system then extracts the team members' messages from their group conversation and transforms them into a structured JSON format representation. This file preserves the key attributes, participants' metadata, and the chronological sequence of team messages. The agent retrieves all messages at scheduled time intervals, triggering data extraction and processing automatically upon the interval’s completion. 

\subsection{Stage\textbf{\textcircled{2}}: Create Feedback Prompts}
\label{Team Dynamics Evaluation}
\toolname{} then evaluates team dynamics using the communication metrics described in Section \ref{prompt-design}. These metrics show how team members interact, collaborate, and contribute to their discussions. The system computes the text-based metrics using pre-coded methods and provides these as input data to the LLM to ensure consistency in analysis and reduce variations in interpretation. A summary of these metrics and their computation is displayed in Table \ref{tab:team_dynamics_evaluation}.

The system calculates the \textit{LSM} scores through a four-step process: (1) The system tokenizes both individual transcripts and the team transcripts to tokens using part-of-speech (POS) tagging to identify specific categories of function words; (2) the system then extracts words from predefined categories (e.g., articles, pronouns) and computes the percentage of function words in each category for both the individual and the team by dividing the number of function words by the total number of words; (3) for each category, it calculates the difference between the percentage of function words of a team member and the percentage of function words of the team; (4) the final LSM score is obtained by averaging these differences across all categories. The LSM score ranges from 0 to 1, where 1 represents perfect alignment between the individual and the team. This result is added to the prompt. The specific implementation details for LSM can be found in Appendix \ref{LSM}.

For \textit{sentiment}, \toolname{} uses the VADER sentiment analysis library \cite{hutto2014vader}. It calculates a compound sentiment score ranging from -1 (negative) to 1 (positive) to represent the overall emotional state of both individual contributions and the entire team discussion. This score is added to the prompt. 

To evaluate \textit{team engagement}, the system tokenizes both the team transcripts and the individual transcripts, then calculates each team member's engagement by dividing the number of words they contributed by the total number of words in the team conversation. To evaluate the overall team engagement, \toolname{} analyzes the complete team conversation along with the total number of team members to determine whether all participants contributed to the discussion. These ratios are then provided to the LLM as input data.

To compute the context-based metrics, the system includes specific evaluation instructions within the prompt and requests the LLM to assess them. For the evaluation of \textit{Transactive Memory System}, the prompt instructs the LLM to examine how effectively team members leveraged each other's expertise, referenced specialized knowledge, and delegated subtasks. To evaluate \textit{collective pronoun} usage, we instructed the LLM to analyze the team transcripts and assess how team members framed their contributions as part of a unified effort using inclusive language. When evaluating \textit{communication flow}, \toolname{} adds both messages' timestamps and team members' usernames to the prompt. Then, the LLM is instructed to evaluate how promptly team members responded to one another and to identify instances of overlapping or interruptive exchanges. Lastly, to assess \textit{topic coherence}, \toolname{} adds the team's defined task goal---embedded in the system by the user---and instructs the LLM to determine how well the team conversation remained focused on the task goal. Details of the prompt are explained in Appendix \ref{apandxPrompts}.

\subsection{Stage\textbf{\textcircled{3}}: LLM Feedback Generation}
\label{LLM Feedback Generation}
After creating the prompts, \toolname{} sends one request to the LLM per team member to generate individual feedback messages and another request to generate team feedback messages. 

\begin{figure*}[!htb]
    \centering
    \includegraphics[width=\linewidth]{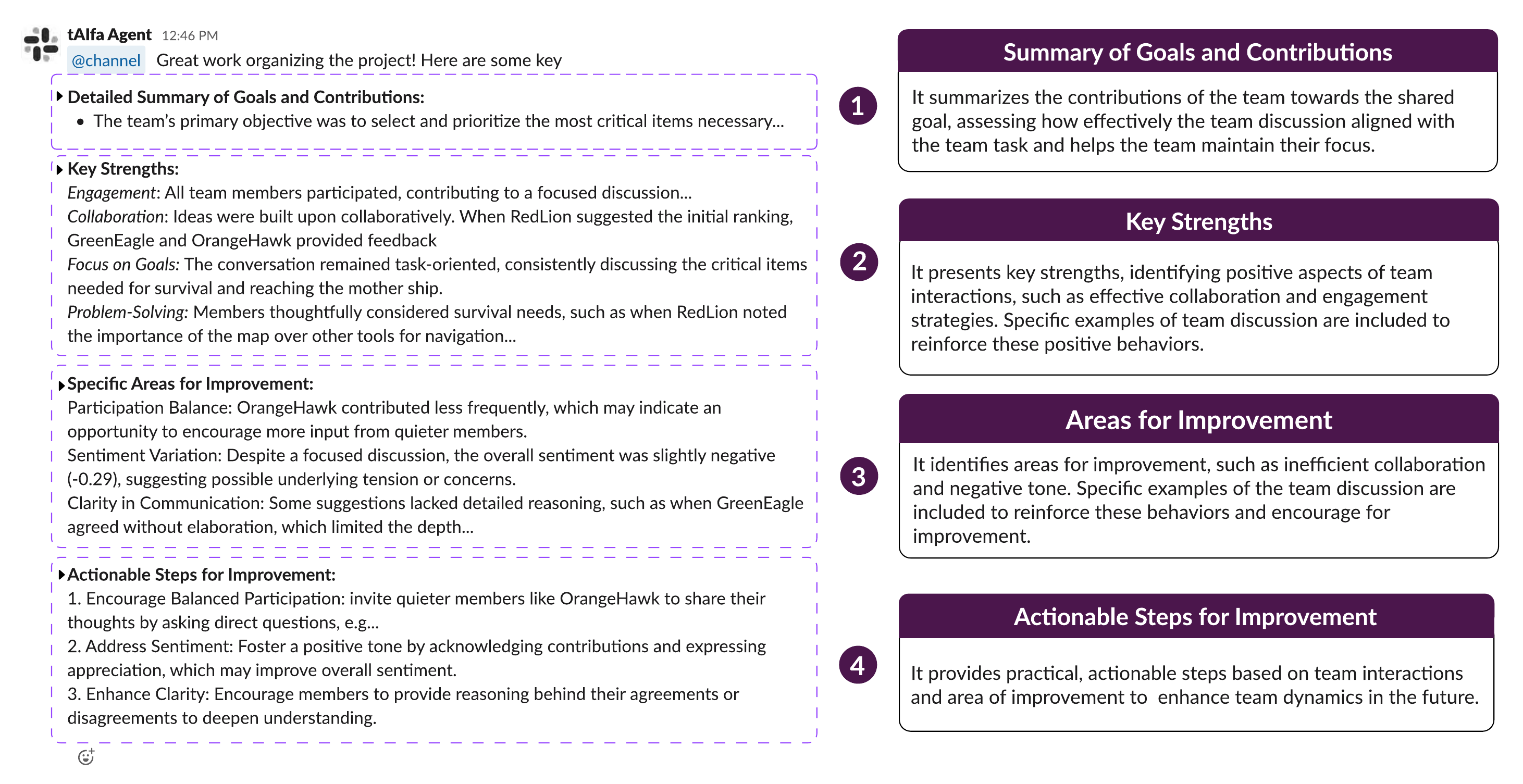}
    \caption{Example of AI-generated team feedback produced by the \toolname{} agent. The feedback is organized into four sections: \textbf{\textcircled{1}} a detailed summary of the team’s goals, \textbf{\textcircled{2}} key team strengths, \textbf{\textcircled{3}} specific areas for improvement, and \textbf{\textcircled{4}} actionable steps.}
    \Description{Example of AI-generated team feedback produced by the \toolname{} agent. The feedback is organized into four sections: \textbf{\textcircled{1}} a detailed summary of the team’s goals, \textbf{\textcircled{2}} key team strengths, \textbf{\textcircled{3}} specific areas for improvement, and \textbf{\textcircled{4}} actionable steps.}
    \label{fig:message-example}
\end{figure*}

\subsection{Stage\textbf{\textcircled{4}}: Deliver Feedback Messages}
\label{Feedback Delevering}
\toolname{} receives the individual- and team-level feedback messages from the LLM and delivers them to the teams. Individual feedback messages are sent as private messages, and the team feedback messages are sent in the group channel. The system organizes the individual feedback messages into four sections (Figure \ref{fig:message-example}). The first one provides a detailed summary of the task goal and the user's contributions in relation to the team's objectives. The second section highlights the key strengths, identifying specific aspects of the user's contributions that positively impacted the conversation and the team. Examples from the discussion are included to highlight these strengths. The third section suggests areas for improvement, offering constructive recommendations on how users can enhance their collaboration with the team. Lastly, the fourth section translates these areas for improvement into actionable steps, providing practical strategies to help users improve their team collaboration. 

Team feedback messages consist of three sections. The first section summarizes the contributions of the team towards the shared goal, assessing how effectively the team discussion aligned with the team task and helping the team maintain its focus. The second section presents key strengths, identifying positive aspects of team interactions, such as effective collaboration and engagement strategies. Specific examples of team discussion are included in the messages to reinforce these positive behaviors. The third section identifies areas for improvement and provides practical, actionable steps based on team interaction to enhance team dynamics.

\subsection{Implementation}  
We developed \toolname{} in Python 3.11 using the Slack Bolt framework\footnote{\url{https://tools.slack.dev/bolt-python/}}. The system uses Microsoft Azure's OpenAI services to employ the \texttt{GPT-o1-preview} model, which has strong reasoning capabilities in complex team interactions, particularly through its ability to handle long prompting \cite{zhong2024evaluation}.

%% file: 06_UserStudy.tex
\section{Main Study}
\label{User Study}
We conducted a between-subjects laboratory experiment to evaluate the impact of \toolname{} on team participation and cohesion. Participants were randomly assigned to teams of three members and asked to resolve three problem-solving tasks within a Slack private channel. This study was preregistered before running the sessions\footnote{\url{https://aspredicted.org/wydq-dpgy.pdf}} and approved by our institution's Institutional Review Board (\#24-02-8358). The experiment aimed to address the following research questions:

\begin{itemize}
    \item \textbf{RQ1:} How does AI-generated automated feedback influence team communication patterns?    
    \item \textbf{RQ2:} How does AI-generated automated feedback affect team dynamics and performance outcomes?
    \item \textbf{RQ3:} How do team members perceive the role and effectiveness of an AI agent providing feedback at individual and team levels?    
\end{itemize}

\begin{figure*}[!htb]
   \centering
   \includegraphics[width=\textwidth]{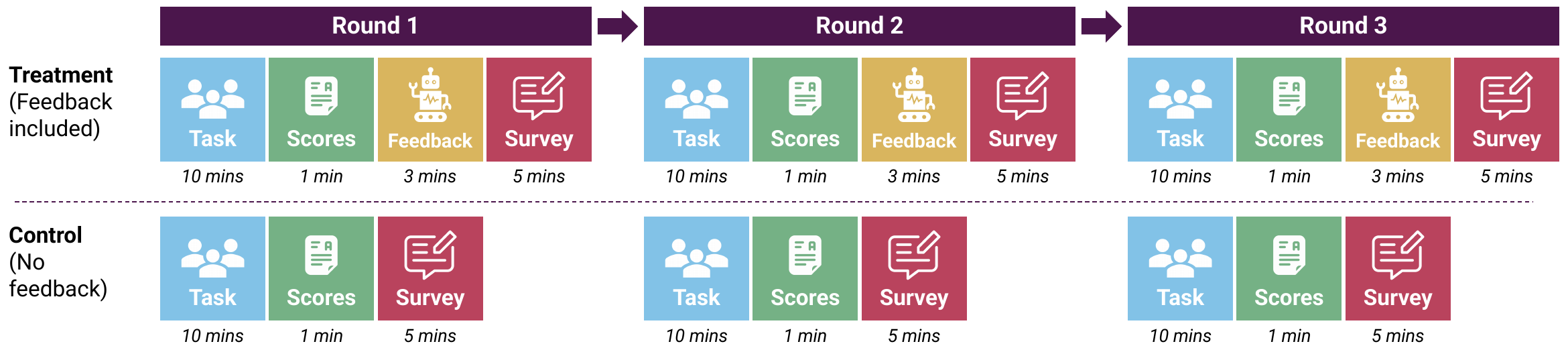}
   \caption{\textit{Main Experiment Outline}. Teams in the treatment condition received LLM-generated feedback from \toolname{} after each task, while teams in the control condition received none. Teams completed three survival tasks and communicated via text on Slack channels. \toolname{} delivered team feedback as public messages and individual feedback as private messages.}
   \Description{Main Experiment Outline. Teams in the treatment condition received LLM-generated feedback from \toolname{} after completing each task, while teams in the control condition received no feedback. The study included three rounds of survival tasks, and teams communicated on Slack channels using text messages. \toolname{} provided team feedback messages as public messages and individual feedback messages as private messages in the channel.}
   \label{fig:user-study}
\end{figure*}

\subsection{Participants}
We recruited 54 participants from two different sources. First, 39 participants signed up through a university-affiliated recruitment system. Most were undergraduate students from a private R1 university in the United States and received extra credit for their participation. Second, we recruited 15 participants via Prolific, an online research platform. We compensated these participants with \$20 USD through the platform. The final participant pool consisted of individuals with varying backgrounds, experience levels, and familiarity with collaborative online tools. All participants reported being proficient in English. 

Each experimental session lasted approximately 60 minutes and was conducted online. They were randomly assigned to teams of three members and in one of two experimental conditions: (1) a control condition with teams working without any automated feedback, or (2) a treatment condition with teams receiving AI-generated feedback by \toolname{}. In total, the study included 18 teams, 10 in the treatment condition and 8 in the control condition.

\subsubsection{Task.}
\label{Task}
We instructed participants to complete three ``survival'' tasks, which require them to imagine themselves in a survival scenario (e.g., being stranded on a desert island) and rank a list of items based on their relevance for survival \cite{nairne2011congruity}. These tasks have been widely used in behavioral studies to examine team collaboration, negotiation, and decision-making processes \cite{smelser2013theory}. We used three adapted versions of this task and evaluated team performance by comparing each team's rankings with the experts' rankings established in previous research. Smaller differences between a team and expert rankings indicate higher presumed expertise. Task instructions and the evaluation formula are shown in Appendix \ref{Experimental_Tasks}.

\subsubsection{Study procedure.}
When signing up for this study, participants reviewed an informed consent form, and only those who consented were included. They then completed a pre-treatment survey to report demographic information and their availability. We scheduled the experimental sessions and invited participants to join via Zoom. Prior to each session, participants were instructed to install the Slack desktop app on their computers. At the beginning of each session, we introduced the experiment's goals and purpose, the tasks to be conducted, and the Slack workspace. This introduction lasted between 5 and 7 minutes. Participants were then given a link to join the Slack workspace, added to the main channel named $\mathtt{\#all\_teams}$, and assigned generic aliases to preserve anonymity. 

\toolname{} created private group Slack channels and randomly assigned participants into teams of three. Each team was then randomly assigned to one of the experimental conditions (See Figure \ref{fig:user-study}). Within their group channels, teams completed three rounds of the survival tasks, with 10 minutes allotted for each round. Participants could chat for the full duration or stop their discussion at any point while waiting for the timer to expire. Once the 10 minutes had passed, the system requested the teams to submit their final item rankings. It then calculated their final scores and displayed the expert rankings and their performance scores. In the treatment condition, participants additionally received AI-generated feedback messages: \toolname{} sent a private message to each member with individual feedback, while it sent the team feedback to the group channel. We gave participants three minutes to read the feedback messages. At the end of each round, all participants completed a post-treatment questionnaire to assess their experience with their teams. In the final round, this survey also included questions regarding the system's usability and perceptions of \toolname{} in the treatment condition. 

\subsection{Measurements}
\subsubsection{Transcripts Metrics.} 
To assess whether \toolname{} affected team communication, we analyzed the transcripts using the following conversation metrics:
\begin{itemize}
    \item \textit{Conversation Duration:} The total time (in minutes) from the first message to the last message, capturing the overall length of team members' interactions. 
    \item \textit{Speaker Turn Frequency:} The number of times that each team member contributed to the discussion, reflecting participation and a balance of interventions.
    \item \textit{Total Word Count:} The total number of words exchanged in the conversation, indicating communication volume.
\end{itemize}

\subsubsection{Task score.}
We measured teams' performance across three rounds of tasks using a percentage score metric, which reflected how closely a team's ranking aligned with the expert ranking relative to the worst possible ranking. To account for differences in task difficulty, scores were normalized per task. Higher percentages indicate greater accuracy (Evaluation formula in Appendix \ref{Evaluation Formula}).

\subsubsection{Self-Reported Measures.}
To check if participants paid attention to the feedback displayed by \toolname{}, we included a manipulation check in the post-task questionnaire. Only two out of 30 participants in the treatment group stated that they did not see any feedback in only one of the rounds.

By the end of each round, participants evaluated in 5-Likert scales their teams' efficiency \cite{brucks2022virtual}, quality of interactions \cite{setlock2004taking}, cohesion \cite{chin1999perceived}, satisfaction \cite{Peeters2006-sw}, and viability \cite{Bayazit2003-ry}. We also use the NASA-TLX 7-point Likert scale to measure participants' task load over three rounds \cite{hart1988development, hart2006nasa}. 

In the final survey, participants evaluated the system's usability using the SUS scale \cite{brooke1996sus} (Cronbach's $\alpha= 0.90$). Participants in the treatment condition evaluated \toolname{}'s overall explainability ($\alpha= 0.78$) and helpfulness---at the individual ($\alpha= 0.96$) and team ($\alpha= 0.96$) level---using the \cite{hoffman2018metrics}'s explainable AI scale. We additionally created a 5-Likert scale with five assertions---such as \textit{``I found the messages provided by the AI Feedback agent more helpful than the ones that a human manager could provide''}---to compare \toolname{}'s feedback to feedback that they have received in the past from managers or mentors (See items in Appendix \ref{human-ai-comparison-item}). The items showed high inter-reliability ($\alpha= 0.81$), and we proceeded to average these items into one single score per participant. Finally, we included open-ended questions to examine participants' perceived benefits and challenges of collaborating within their teams and using \toolname{}.

\subsection{Analyses}
We first conducted \textit{t}-tests on the quantitative measures to assess differences between the control and treatment teams by the end of the third round. This analysis included transcript-based metrics, task scores, and self-reported measures. To evaluate the normality of each metric, we applied the Shapiro–Wilk test and used a Box-Cox transformation for variables that violated the normality assumption. We then employed Welch's \textit{t}-test to compare both conditions, accounting for unequal sample sizes between conditions, and report Cohen's \textit{d} for effect sizes.

To examine changes in the variable over time and assess the effect of \toolname{}, we conducted repeated measures ANOVA with time (i.e., three tasks) as the within-subject factor and condition (i.e., treatment vs. control) as the between-subjects factor. 

To analyze participants' responses, two researchers independently inductively coded the data, then they met to cross-check and validate their codes. Since the participants responded to questions that asked specifically about the advantages and obstacles of working with \toolname{}, the researchers coded the data with this context in mind. A third author joined the coding discussion when necessary. The inter-rater agreement, measured using Cohen's kappa, was 0.90, indicating high agreement. From the discussion, the researchers surfaced several themes from the analysis. We describe the main emerging topics in Section \ref{sec:qual_analysis}.

%% file: 07_Results.tex
\section{Results}
In this section, we present the findings from our analysis of the team conversation metrics and quantitative survey data, followed by the results of qualitative survey analysis.

\subsection{Quantitative Analysis}
\begin{figure*}[!htbp]
  \centering
  \begin{subfigure}[b]{0.33\textwidth}
    \includegraphics[width=\textwidth]{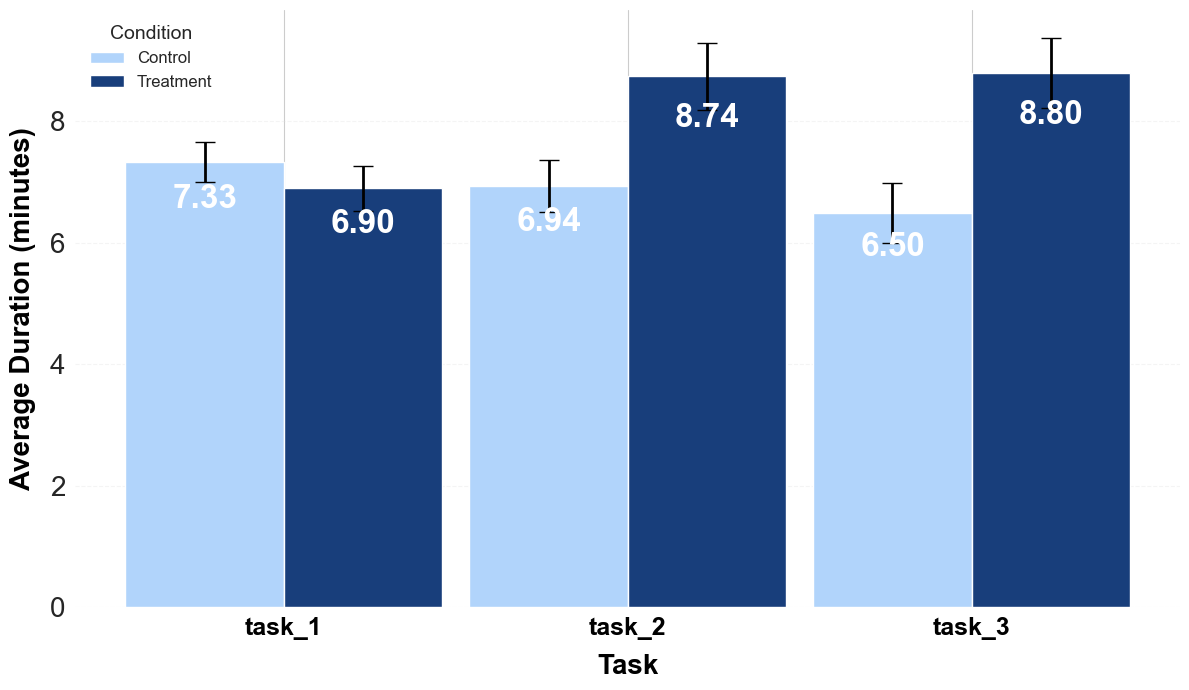}
    \caption{Average Conversation Duration}
    \label{fig:conversation_duration}
  \end{subfigure}
  \hfill
  \begin{subfigure}[b]{0.33\textwidth}
    \includegraphics[width=\textwidth]{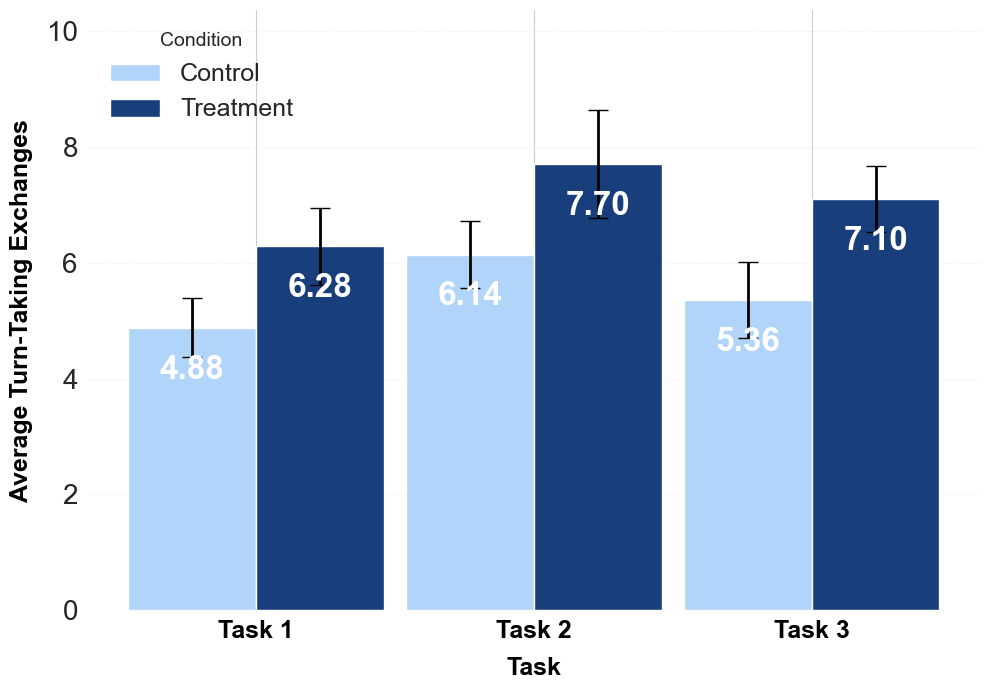}
    \caption{Average Speaker Turn}
    \label{fig:turn_taking}
  \end{subfigure}
  \hfill
  \begin{subfigure}[b]{0.33\textwidth}
    \includegraphics[width=\textwidth]{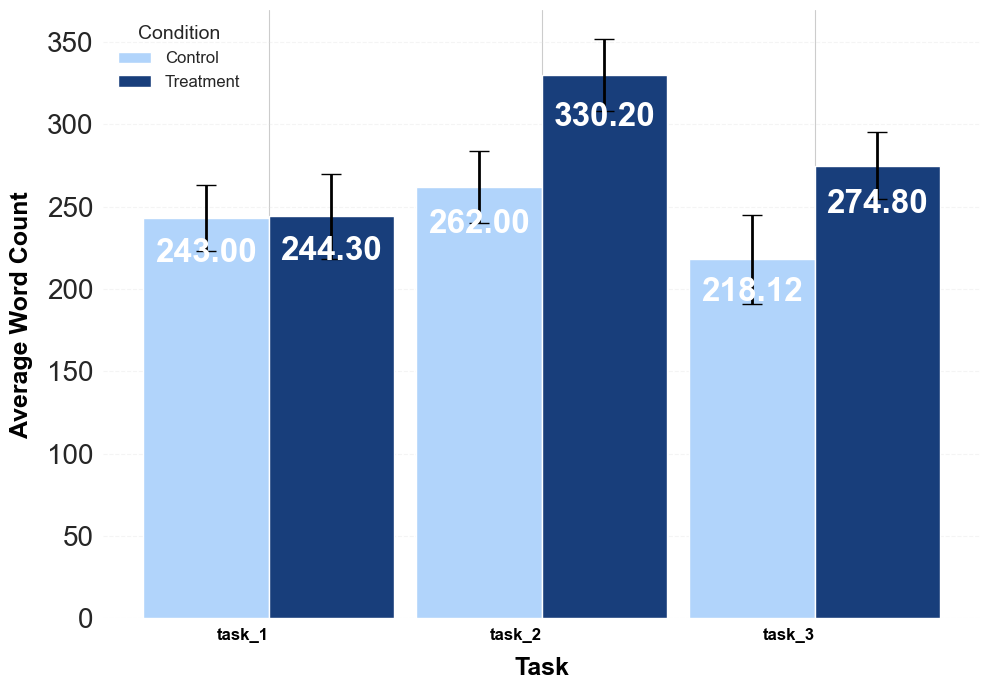}
    \caption{Average Word Count}
    \label{fig:average_word_count}
  \end{subfigure}
  \vspace{0.2em} 
  \begin{subfigure}[b]{0.33\textwidth}
    \includegraphics[width=\textwidth]{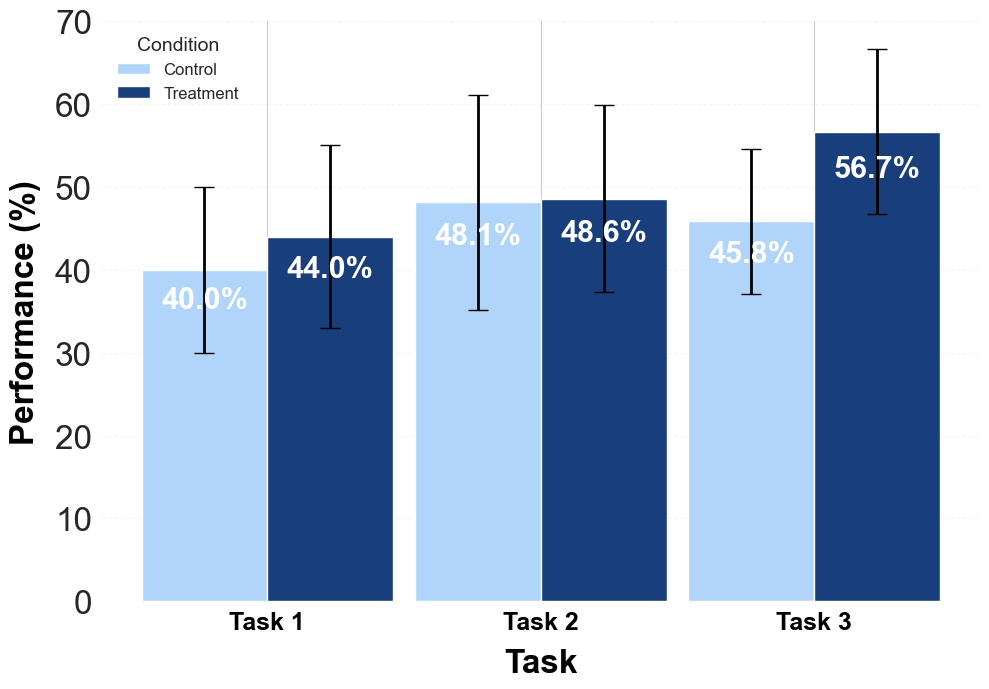}
    \caption{Task Performance}
    \label{fig:task_performance}
  \end{subfigure}
  \hspace{1em}
  \begin{subfigure}[b]{0.33\textwidth}
    \includegraphics[width=\textwidth]{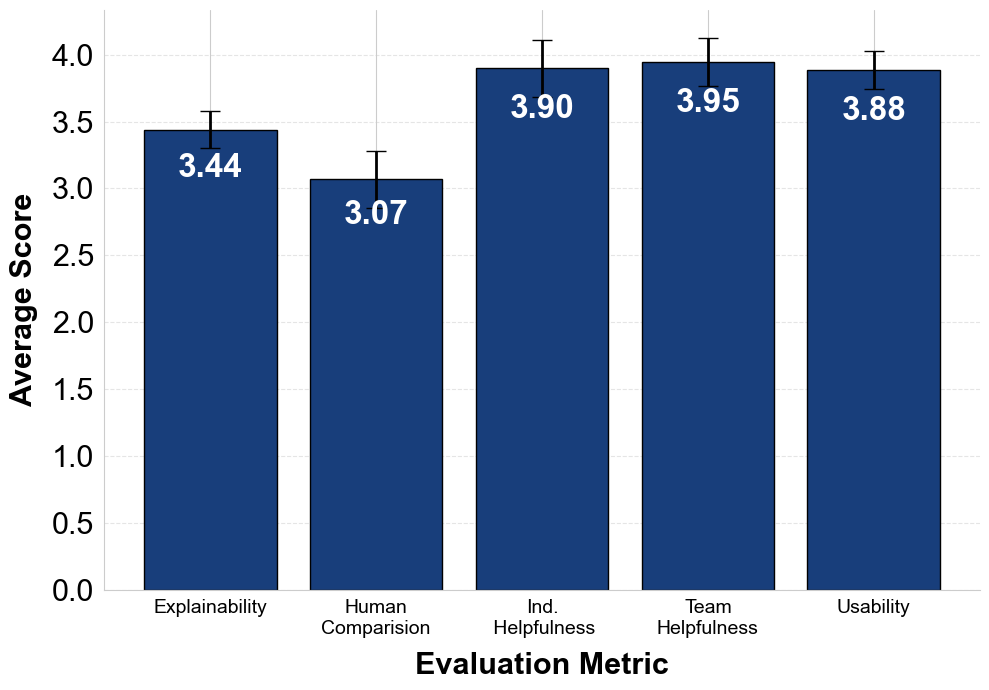}
    \caption{User Perceptions of AI‐generated feedback.}
    \label{fig:usability}
  \end{subfigure}
  \caption{Team communication metrics and task performance between conditions. Error bars denote standard errors.}
  \label{fig:combined_figures}
\end{figure*}

\paragraph{Conversation Duration.} The duration of team discussions varied notably across tasks (Figure~\ref{fig:conversation_duration}). While most teams in the control condition did not use the full 10 minutes allotted per task, several teams in the treatment condition continued chatting until the time expired. In Task 1, the control teams spent an average of 7.33 minutes ($SD=1.60$) discussing, compared to 6.90 minutes ($SD=2.04$) for the treatment teams. In Task 2, the treatment teams ($M=8.74, SD=2.99$) chatted 27\% more than the control teams ($M=6.94, SD=2.10$). Lastly, in Task 3, the treatment teams ($M=8.80, SD=3.11$) discussed 35\% more than the control teams ($M=6.50, SD=2.41$). The difference in Task 3 was statistically significant ($t(49.35)=-2.11,p<0.05$). A repeated measures ANOVA revealed a significant main effect of the treatment condition ($F(1, 154) = 4.21, p < 0.05, d=0.57$), indicating that the duration of team conversations was different across conditions. However, the interaction term was not significant, suggesting that groups in both conditions followed a similar pattern of change over time.

\paragraph{Speaker Turn Frequency.}
We found that \toolname{} enabled more participants to contribute to their conversations (Figure~\ref{fig:turn_taking}). In Task 1, participants in the treatment teams had, on average, 6.28 exchanges ($SD=3.65$), while control teams' members had, on average, 4.88 exchanges ($SD = 2.47$). In Task 2, treatment teams' members ($M=7.70, SD=5.11$) had 25\% more exchanges than the control teams' members ($M=6.14, SD=2.83$). Lastly, in Task 3, the treatment teams' members ($M=7.10, SD=3.10$) had 32\% more turn-taking exchanges than the control teams' members ($M=5.36, SD=3.21$). This difference was  significant ($t(50.52)=-2.03,p<0.05,d=0.55$). A repeated measures ANOVA revealed a significant main effect of the condition ($F(1, 154) = 5.74, p < 0.05$), but the interaction term was not significant.

\paragraph{Total Word Count.}
On average, teams in the treatment condition used more words in each task than those in the control condition (Figure~\ref{fig:average_word_count}). In Task 1, treatment teams had a similar number of words ($M = 244.30, SD = 141.75$) compared to the control teams ($M = 243.00, SD = 97.81$). In Task 2, treatment teams used 26\% more words ($M = 330.20, SD = 119.20$) than control teams ($M = 262.00, SD = 107.73$). We observed the same difference in Task 3, where the treatment teams also used 26\% more words in their conversations ($M = 274.80, SD = 111.39$) than the control teams ($M = 218.12, SD = 132.99$). However, this difference was not statistically significant in Task 3 (Welch Two Sample t-test, $t(49.74)=-1.56, p > 0.05$). A repeated measures ANOVA also showed that the treatment condition was not a significant factor ($F(1, 154) = 2.78, p > 0.05$).

\paragraph{Team Performance.} 
On average, teams in the treatment condition achieved slightly higher performance scores than those in the control condition (Figure \ref{fig:task_performance}). In Task 1, treatment teams scored an average of 44\% ($SD = 0.35$), compared to 40\% ($SD = 0.28$) of the teams in the control condition. In Task 2, treatment teams scored 48.6\% ($SD = 0.36$) versus 48.1\% ($SD = 0.37$) for control teams. Only in the last round, Task 3, we saw a larger difference between the treatment teams, scoring on average 56.7\% ($SD = 0.31$), and control teams, which scored on average 45.8\% ($SD = 0.24$). However, this difference was not statistically significant ($t(15.80)=-0.93, p>0.05$).

\begin{figure*}[!htb]
   \centering
   \includegraphics[width=\textwidth]{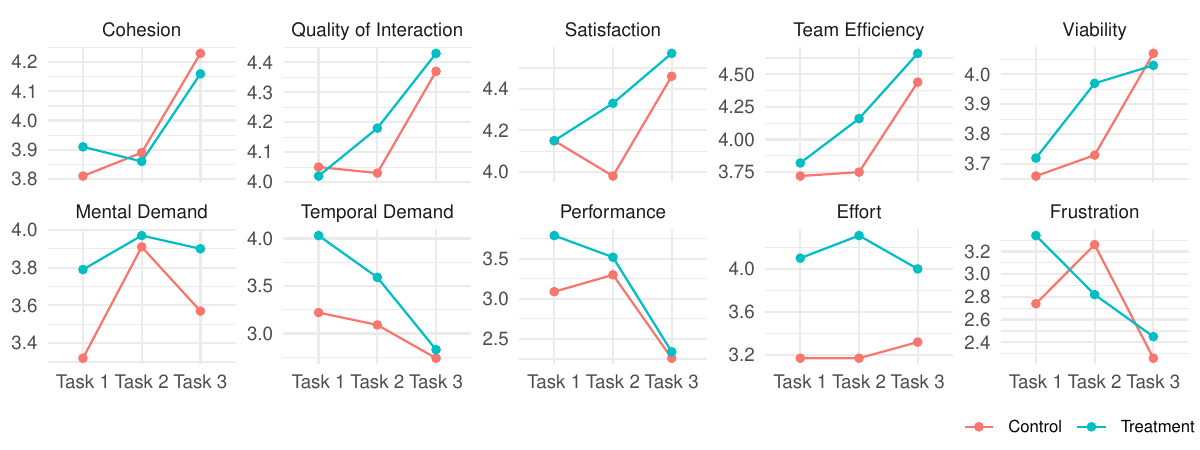}
   \caption{Participants' survey responses.}
   \Description{}
   \label{fig:survey-evolution}
\end{figure*}

\paragraph{Survey Metrics.}
Table~\ref{tab:team-dynamics} presents the survey results after the final task, while Figure \ref{fig:survey-evolution} shows the average scores across all tasks. By the final round, participants in the treatment condition reported slightly higher levels of team efficiency ($M=4.66, SD=0.59$) than participants in the control condition ($M=4.44, SD=0.77$). They also rated the quality of their team interactions slightly higher ($M=4.43,SD=0.71$) than control participants ($M=4.37, SD=0.48$). However, the participants who used \toolname{} reported marginally lower scores for team satisfaction ($M_{T}=4.46, SD_{T}=0.72$ vs. $M_{C}=4.57, SD_{C}=0.89$), lower levels of team cohesion ($M_{T}=4.16, SD_{T}=0.90$ vs. $M_{C}=4.23, SD_{C}=0.62$), and lower agreement on working with the same team again ($M_{T}=4.03, SD_{T}=0.98$ vs. $M_{C}=4.07, SD_{C}=0.71$) than the control participants. None of these differences between conditions was statistically significant.

Regarding the NASA TLX measures, the treatment participants reported higher scores than the control participants across all the items. We found that effort (i.e. \textit{``How hard did you have to work to accomplish your level of performance?''}) to be considerably higher in the treatment condition ($M=4.00, SD=1.71$) compared to the control condition ($M=3.32,SD=1.25$). Although we did not find any significant differences between both conditions after the final round ($t(47.61)=-1.47,p>0.05$), a two-way repeated measures ANOVA shows that the condition was a significant factor ($F(1,149)=15.36, p < 0.05$).

\begin{table}[!htb]
\renewcommand{\arraystretch}{0.8}
\begin{tabular}{@{}lll@{}}
\toprule
\textbf{Metric} & \textbf{Control} \ & \textbf{Treatment} \\ \midrule
Team Efficiency & $4.44 \pm 0.77$ & $4.66 \pm 0.59$ \\ 
Quality of Interaction & $4.37 \pm 0.48$& $4.43 \pm 0.71$\\ 
Cohesion & $4.23 \pm 0.62$ & $ 4.16 \pm 0.90$\\ 
Satisfaction & $4.57 \pm 0.89$ & $4.46 \pm 0.72$\\ 
Viability & $4.07 \pm 0.71$& $4.03 \pm 0.98$\\ \midrule
Mental Demand & $3.57 \pm 1.41$ & $3.90 \pm 1.59$\\ 
Temporal Demand & $2.74 \pm 1.29$ & $2.83 \pm 1.67$\\ 
Performance & $2.26 \pm 1.36$ & $2.34 \pm 1.63$\\ 
Effort & $3.32 \pm 1.25$ & $4.00 \pm 1.71$\\ 
Frustration & $2.26 \pm 1.18$ & $2.45 \pm 1.53$\\ \bottomrule
\bottomrule
\end{tabular}
\caption{Team dynamics scores (5-Likert scale) and NASA TLX scores (7-Likert scale) after the final round. Means and standard deviations are reported. \# of observations: 52 users.}
\label{tab:team-dynamics}
\end{table}

Figure~\ref{fig:usability} presents the average participant ratings of \toolname{} across five evaluation metrics. Participants rated the helpfulness of the team feedback messages the highest ($M=3.95,SD=0.98$), closely followed by individual feedback messages' helpfulness ($M=3.90, SD=1.15$) and usability ($M=3.88,SD=0.77$), suggesting that \toolname{}'s messages and ease of use were perceived positively. Explainability received a moderate score ($M=3.44,SD=0.76$), indicating that while participants found the messages understandable, there remains room for improvement. The lowest-rated metric was comparing \toolname{} to feedback provided by mentors or leaders ($M=3.07, SD=0.97$), suggesting participants were less convinced that the LLM agent matched the quality of feedback typically offered by a human manager. 

\subsection{Qualitative Analysis}
\label{sec:qual_analysis}
Overall, participants in the treatment condition had positive sentiments about their experience working with \toolname{}. Despite prompting for both positive and negative feedback, researchers found noticeably more positive comments than negative ones. We summarize the main topics identified in their comments.

\subsubsection{Perceived Benefits of AI-Generated Feedback.}
The majority of participants noted that receiving timely feedback was a significant advantage of the \toolname{}. Participants noted that the feedback \textit{``... was fast and I liked the feedback it provided''} (P7), and \textit{``... was rather fast given the amount of information it had to process''} (P8). In addition, P20 said that the feedback provided \textit{``... was very detailed in highlighting what we really needed to improve and got us going''}. Participants specifically liked the statistical information provided in the AI-generated feedback and the objectivity of the \toolname{}.

The AI helped participants reflect on areas for improvement by providing actionable feedback. P28 noted that the feedback \textit{``gave me some tips to think about.''} Other participants noted that feedback \textit{``helped us to know what to do better''} (P19) as well as telling participants \textit{``when to improve''} (P22). 

\subsubsection{Perceived Challenges of AI-Generated Feedback.}
Many participants noted \toolname{}'s lack of understanding of contextual statements and slang as an obstacle. Although the LLM could read the whole conversation transcript, there were still issues in connecting context across several messages, especially those requiring knowledge of popular culture or text slang. Participants stated that \textit{``...sometimes it did not take the overall chat into context ... focusing on a single comment and not linking it to another''} (P19), or that \toolname{} was \textit{``not understanding jargon that we all understand''} (P1). P7 \textit{``noticed each time there was about 20 percent of information he [the agent] provided [sic] was inaccurate''}.

Other participants disliked the lack of human touch, such as P18, who said that \toolname{} \textit{``.. is missing human touch and empathy.''} Many participants felt the feedback was \textit{``impersonal''} (P26), \textit{``monotonous''} (P27), and \textit{``wasn't always the kindest''} (P22). Interestingly, P15 noted this as a positive characteristic, saying \textit{``\toolname{} does not have any emotional filter, so I was able to hear the absolute truth, unlike how a human would `tiptoe' around others' feelings.''}

%% file: 08_Discussion.tex
\section{Discussion}
In this study, we designed and tested an LLM-based system called \toolname{} that generates automated feedback for both individuals and teams based on their interactions and discussions. Our work sheds light on how advanced AI technologies, such as LLMs, can become instrumental partners in enhancing team communications and dynamics. We discuss how these findings provide insights into the role of AI in facilitating feedback and supporting collaboration. 

\subsection{RQ1: Automated feedback impacted teams' communication patterns}
Our findings indicate that teams receiving feedback from \toolname{} had talked more and contributed more evenly to the conversation. Employing communication metrics for \toolname{}'s feedback analysis helped participants become more aware of their interaction patterns and individual contributions. These behavioral shifts align with previous research illustrating that feedback can enhance awareness of team members' participation \cite{bimba2017adaptive, kinder2025effects}. For example, \toolname{}'s interventions identified imbalances in participation, encouraged quieter members to contribute more, and prompted more vocal participants to practice mindful listening. These interventions fostered longer and more balanced discussions. 

The increase in both the duration and participation in team conversations allowed teams to better engage in the task by integrating diverse knowledge and perspectives \cite{engel2015collective}. In our study, the observed increase in conversation duration and participation within the treatment condition suggests that \toolname{} supported teams in acknowledging each other's ideas and collaboratively exploring the problem space. The \toolname{} agent also promoted better practices among team members, acting as a motivation for more active and engaged conversations. 

Participants' increased engagement was reflected in the significant rise in conversation duration following the delivery of the first feedback session. During the first task, both groups exhibited a relatively similar duration of conversation; however, after receiving feedback from \toolname{}, treatment teams showed a significant increase in length of conversation time, whereas the control group remained relatively stable. The increased time of discussion following the first session suggests that receiving shared feedback may have prompted team members to engage more actively with one another. One explanation is that \toolname{} provided clear and actionable feedback into their interaction, making each participant's contribution more explicit. Consequently, team members likely became more conscious of their interaction, dedicating more time to clarifying, discussing, and reflecting on each other's contributions. 

By design, \toolname{} delivered both team and individual feedback messages focused on communication metrics, which likely served as a conversational ice-breaker that gave members a shared reference point for collective reflection. For example, in several treatment sessions, we observed participants referencing \toolname{}'s feedback from the first task before starting the next one, using it to initiate discussion or clarify expectations. This aligns with Tuckman's theory of group development \cite{tuckman1965developmental}, which suggests that teams in the early forming stage often rely on the shared topic to learn about each other and deepen collaboration. Given this perspective, receiving the first round of feedback from \toolname{} likely helped teams gradually develop stronger communication patterns over time \cite{hohenstein2023artificial, muresan2024should}. 

\subsection{RQ2: The lack of time and task-specific guidance constrained team improvements}
While \toolname{} positively influenced team communication behaviors, it did not lead to significant improvements in team performance. Differences between treatment and control groups in performance or team dynamics were relatively small. One possible explanation for these results is that \toolname{} strongly emphasized communication patterns in its feedback rather than offering actionable guidance tied to task success (e.g., \textit{``You may benefit from categorizing the items (e.g., navigation, protection, sustenance) and then ranking within those groups.''}. \toolname{}'s prompts focused on language use, understanding of each other's skills and ideas, and participation, but did not assess which members' actions could have improved, whether they made mistakes, or how well-connected and satisfied they felt within the team. Neither did it provide specific strategies nor instructions for solving the task more effectively. Future versions could instruct the LLM to infer what type of task or actions participants are working on and offer task-oriented feedback, potentially requiring additional context input. 

Moreover, the short duration of our study may have also limited the teams' ability to meaningfully incorporate feedback into their collaborative behavior. Behavioral change in teams often requires time for reflection, repeated interactions, and the gradual norm development \cite{otte2018effective, schippers2013reflect}. Moreover, the online and episodic nature of the study introduces additional challenges. Unlike most real-world teams, participants did not know each other before the task and lacked sustained interaction. Feedback might also be more effective if delivered through richer communication channels, such as phone/video calls or in-person meetings, as body language can also promote connections and awareness between members \cite{ishii2019revisiting}. Furthermore, time pressure to complete the task may have disrupted team dynamics, negatively affecting team members' engagement \cite{kerr2004group}. Consequently, while participants responded positively to \toolname{}'s feedback, providing task-specific suggestions and allowing more time could have generated significant improvements in team dynamics. 

\subsection{RQ3: Positive user perceptions of the AI agents in the team}
While \toolname{} could generate realistic and eloquent feedback messages, most participants were neutral about evaluating this LLM agent better than a human manager. The qualitative responses offered more details that highlight the potential benefits and challenges of AI-generated feedback. Many participants valued the speed and objectivity of the feedback generated by \toolname{} compared to human feedback, who are often delayed or unavailable. For example, P7 stated that receiving feedback quickly allowed them to reflect and understand where to grow. Similarly, P15 also cited the helpfulness of receiving quick feedback and viewed the AI's lack of an emotional filter as a strength that enhanced its objectivity compared to human input. However, others saw this same lack of emotion as a limitation. Participants P22 and P27 described the feedback as ``monotonous'' or ``unkind,'' expressing a preference for the empathy and contextual understanding that human managers can provide. These varied reactions to AI feedback align with prior research showing individual differences in preferences for feedback tone and directness \cite{okoso2025expressions}.

Beyond tone and emotional aspects, participants also reflected on the clarity and accuracy of \toolname{}'s feedback. Most participants rated \toolname{}'s feedback as clear and explainable, yet others perceived that \toolname{} sometimes misinterpreted slang or informal language (e.g., P1, P7, P19). While attitudes toward AI are subjective and dependent on external factors and personal experience \cite{grassini2023hope}, much of users' trust in AI can be tied to their perception of the system's competency \cite{krop2024effects}. In addition to perceptions of quality and trust, participants also reflected on the cognitive effort required to engage with \toolname{}'s feedback during the tasks. Participants reported that \toolname{} imposed additional effort, likely due to the added step of reading and interpreting the feedback while under time pressure. However, temporal demand converged with the control condition by Task 3, suggesting that teams became faster at processing \toolname{}’s feedback after brief exposure.

Based on participants' experiences, our findings point to several actionable design directions for future versions of \toolname{}. One is that contextual grounding of the agent can reduce misinterpretations and impersonally perceived incompetence. While we intentionally avoided using users' personal data for feedback generation, future versions of \toolname{} could offer more personalized, adaptive feedback by incorporating users' prior interactions, performance, personality traits, and preferences \cite{bimba2017adaptive}. Thus, \toolname{} could frame feedback messages according to user preferences (e.g., straightforward or detailed). Allowing \toolname{} to employ behavioral‑learning competency could further mitigate impersonality while preserving objectivity. 

Another way to improve \toolname{}'s situational awareness and relevance in group interactions would be to incorporate memory across multiple conversations to maintain richer contextual information about team members \cite{Jiang28052023}. Consequently, \toolname{} would consider users' expertise: novice users or newly formed groups could receive lay instructions, whereas skilled users may prefer more specific and detailed feedback \cite{kinder2025effects,kalyuga2007expertise}. By aligning feedback with users' expectations and needs, \toolname{} could be perceived as more helpful and relevant to their context.

\subsection{Adopting and Customizing \toolname{} in the Workplace}
We designed \toolname{}'s framework to seamlessly integrate with existing communication platforms, such as Slack, Discord, or Microsoft Teams. \toolname{} can be developed in these platforms with the support of lightweight agents connected to production servers and LLMs. Moreover, other and more recent LLMs can be employed in this application. By leveraging this digital infrastructure, \toolname{} can be a flexible and interoperable solution that teams can install within their online workspaces without altering their current workflows. To support the administration of this system, \toolname{} could also provide a dashboard that allows administrators to customize the feedback prompts, modify instructions, and adjust the system's actions to the specific needs and dynamics of the teams. 

While \toolname{} can help workers improve their communication interactions, it can assist workplace administrators, team leaders, and managers to scale and provide feedback more frequently. A potential case scenario for \toolname{} is in helping newly-formed remote teams, which need special guidance to effectively collaborate given their fragmented communication and silos \cite{yang2022effects}. For example, team leaders can add \toolname{} to their project channels and specify time frames for the system to provide periodic feedback, accounting for time zone differences or work schedules. Also, team leaders can specify guidance that \toolname{} should include to provide customized feedback to the team members. Importantly, in these early stages, \toolname{} can act as a conversational icebreaker, helping team members initiate dialogue and build trust. Administrators can further customize \toolname{} to align with their workplace needs, including configuring feedback message prompts to reflect the organization's norms and priorities. Administrators might predefine certain feedback templates to address recurring challenges---such as improving task delegation or fostering collaboration during brainstorming sessions---as well as for specific roles within the teams. With specific fine-tuning and external support, \toolname{} could learn from employees' previous actions and experiences to provide more customizable feedback. 

\subsection{Limitations and Future Work} 
Our work is not exempt from limitations. First, we did not enable any interactions between \toolname{} and the users, which limited the agent's capacity to address user queries or engage in ongoing dialogue. In designing \toolname{}, we considered the tradeoff between providing real-time support and the risk of overwhelming users with excessive interaction. A future version should consider enabling \toolname{} to interact and dialog with users, answer questions, initiate conversations, and proactively offer suggestions when communication challenges are detected. 

Another limitation was the nature of the task used in our main study. Although survival scenario tasks are structured and widely employed in team experiments, real-world teams rarely interact with this type of task. Future work should evaluate \toolname{} with a wider range of tasks, including ongoing activities in organizations and industries (e.g., prepare a presentation, prepare a report). Expanding the scope of tasks will enhance the system's generalizability in different contexts. Additionally, \toolname{} gave teams their feedback only at the end of the task, missing opportunities to guide users during the process of team collaboration. Future iterations should consider providing real-time feedback---which could be task-specific---and determining an optimal balance between immediacy and helpfulness. 

Moreover, while anonymizing participants' names helped protect their confidentiality, it may have affected how participants perceived \toolname{}'s feedback. Addressing participants with nicknames rather than their real names could have made the feedback less personal. Future work should test the effect of AI-generated feedback on non-anonymized participants to mitigate the effects of our experimental designs. Similarly, relying only on communication metrics may overlook important cognitive and behavioral aspects of teamwork (e.g., thinking, awareness, quality of contributions). Future versions of \toolname{} should aim to promote and assess a broader range of cognitive behaviors by incorporating additional techniques and evaluation measures.

Finally, the participant sample in the main study was relatively small and may not fully represent the broader population. Moreover, a large portion of participants were undergraduate students, which limits the applicability of the findings to professional contexts. Future work should include larger samples to enable more robust statistical analyses, recruit participants active in the workforce to enhance the ecological validity and the generalizability of the findings, and evaluate \toolname{} deployed within real-world organizations for prolonged periods.

%% file: 09_Conclusion.tex
\section{Conclusion}
In this work, we explore the benefits and challenges of providing constructive feedback for teams collaborating in online environments. Building on previous research in team dynamics and conversation patterns, we designed \toolname{}, an LLM-based agent that provides timely, personalized feedback aimed at enhancing team participation and cohesion. We conducted a user study with 18 teams performing problem-solving tasks on Slack, showing that \toolname{} contributed to more balanced and active conversations. By offering feedback messages for individuals and group messages to teams, the system helped align their efforts toward the primary task goals. We envision \toolname{} as a digital tool that can be widely adopted and further explored by researchers and practitioners to support effective teamwork and learning in online work settings.

%% file: 10_EthicsStatement.tex
\section{Ethics statement}
 Both studies were reviewed and approved by Notre Dame's Institutional Review Board ({\#24-02-8358}). Participants were informed of their rights to participate or withdraw from the study at any time during the formative study and the main user study. All participants' names were anonymized by assigning random aliases on Slack. Interactions with \toolname{} took place within the private channels of the Slack workplace. Ethical considerations relevant to the use of LLM models in providing individual and team feedback messages were taken into account. Anonymized participant names, participant titles, and messages were the only information transferred from Slack channels to the LLM model, while other user data were removed.

%% file: Appendix/FormativeStudyPrompts.tex
\section*{APPENDIX}
\section{Formative Study}
\label{appendix: Formative Study Prompts}
\renewcommand\thefigure{\thesection.\arabic{figure}}    
\renewcommand\thetable{\thesection.\arabic{table}}    
\setcounter{figure}{0}    
\setcounter{table}{0}   

\subsection{Corpus Description}
\label{Corpus Descrption}
In our formative study, the discussion of the team was derived from the Group Affect and Performance Corpus (GAP) \cite{braley2018gap}. We utilized a specific conversation dataset titled \texttt{CSV\_Group20\_Oct\_22\_135}\footnote{\url{https://convokit.cornell.edu/documentation/gap.html}}. This data set involved three participants engaged in a six-minute problem-solving task, during which they were required to rank several items according to their importance, from highest to lowest.


\subsection{Feedback Evaluation Criteria}
These evaluation criteria were adapted based on the metrics for explainable AI discussed by Hoffman et al. \cite{hoffman2018metrics}. Table \ref{tab:evaluation_criteria} provides the specific questions and explanations used in our formative study.

\begin{table}[!htb]
\centering
\caption{Nine evaluation criteria used to assess different aspects of nine different candidates' AI-generated feedback prompts. Each criterion is presented as a question, accompanied by a brief explanatory note to help participants understand how to interpret and apply the criterion when rating the feedback they received.}
\label{tab:evaluation_criteria}
\resizebox{\columnwidth}{!}{
\begin{tabular}{p{2.5cm} p{9.5cm}}
\toprule
\textbf{Evaluation Criteria} & \textbf{Question and Explanation} \\ 
\midrule
 Clarity and Understanding & The feedback was clear, concise, and easy to understand. \\
& \emph{Evaluate whether the feedback was written in a straightforward and concise manner. Did the wording make the feedback simple to interpret, or were there parts that seemed confusing or unclear?} \\[0.5em]
\midrule

Feedback Message Satisfaction & The feedback effectively addressed strengths and areas for improvement to enhance performance in future tasks based on my performance. \\
& \emph{This statement asks if the feedback provided addressed the strengths and areas where you can improve. Did it meet your expectations and give you useful insights that make you feel satisfied with the feedback you received?} \\[0.5em]
\midrule

Level of Detail & The feedback provided me sufficient detail to help you improve my performance. \\
& \emph{This statement asks whether the feedback contained enough information for you to understand how to improve. Did it give you the right amount of detail to help you take action, or did it feel too vague or overwhelming?} \\[0.5em]
\midrule

Comprehensiveness & feedback covered all the relevant and important aspects of the task that are necessary to improve in similar future tasks. \\
& \emph{This statement is about whether the feedback covered all the important aspects of your performance. Do you feel like nothing was left out? Good feedback should be thorough and address everything relevant to your task.} \\[0.5em]
\midrule

Accuracy and Reliability & The feedback accurately reflected my strengths and weaknesses. \\
& \emph{You're being asked to evaluate whether the feedback correctly reflected your performance. Did it accurately highlight your strengths and weaknesses? Also, think about whether the feedback was consistent with what you experienced during the task. Reliable feedback should feel truthful and valid.} \\[0.5em]
\midrule

Actionable Feedback & The feedback provided clear and practical steps for improvement. \\
& \emph{This statement focuses on whether the feedback gave you specific and practical steps you can take to improve. Consider whether the feedback gave you clear instructions or ideas you can use to make progress.} \\[0.5em]
\midrule

Task Relevance & The feedback aligned well with the goals of the task. \\
& \emph{Here, you're being asked if the feedback aligned with the task's goals. Did the feedback help you understand how well you performed in relation to what the task was trying to achieve? Good feedback should directly connect to the task's objectives.} \\[0.5em]
\midrule

Engagement & The feedback maintained the recipient's attention without becoming tedious or repetitive. \\
& \emph{Engagement measures how well the feedback message holds the recipient's attention and maintains their interest. A highly engaging message captivates the reader by being dynamic, relevant, and easy to follow, avoiding unnecessary complexity or monotony.} \\[0.5em]
\midrule

Cognitive Efforts & The feedback message was too long and required significant effort to understand. \\
& \emph{Consider whether the feedback required significant mental effort. Was it presented in a way that was intuitive and actionable, or did it feel challenging or overly complex to grasp?} \\[0.5em]

\bottomrule
\end{tabular}}
\end{table}

\begin{figure*}[!htb]
  \centering
  \includegraphics[width=\textwidth]{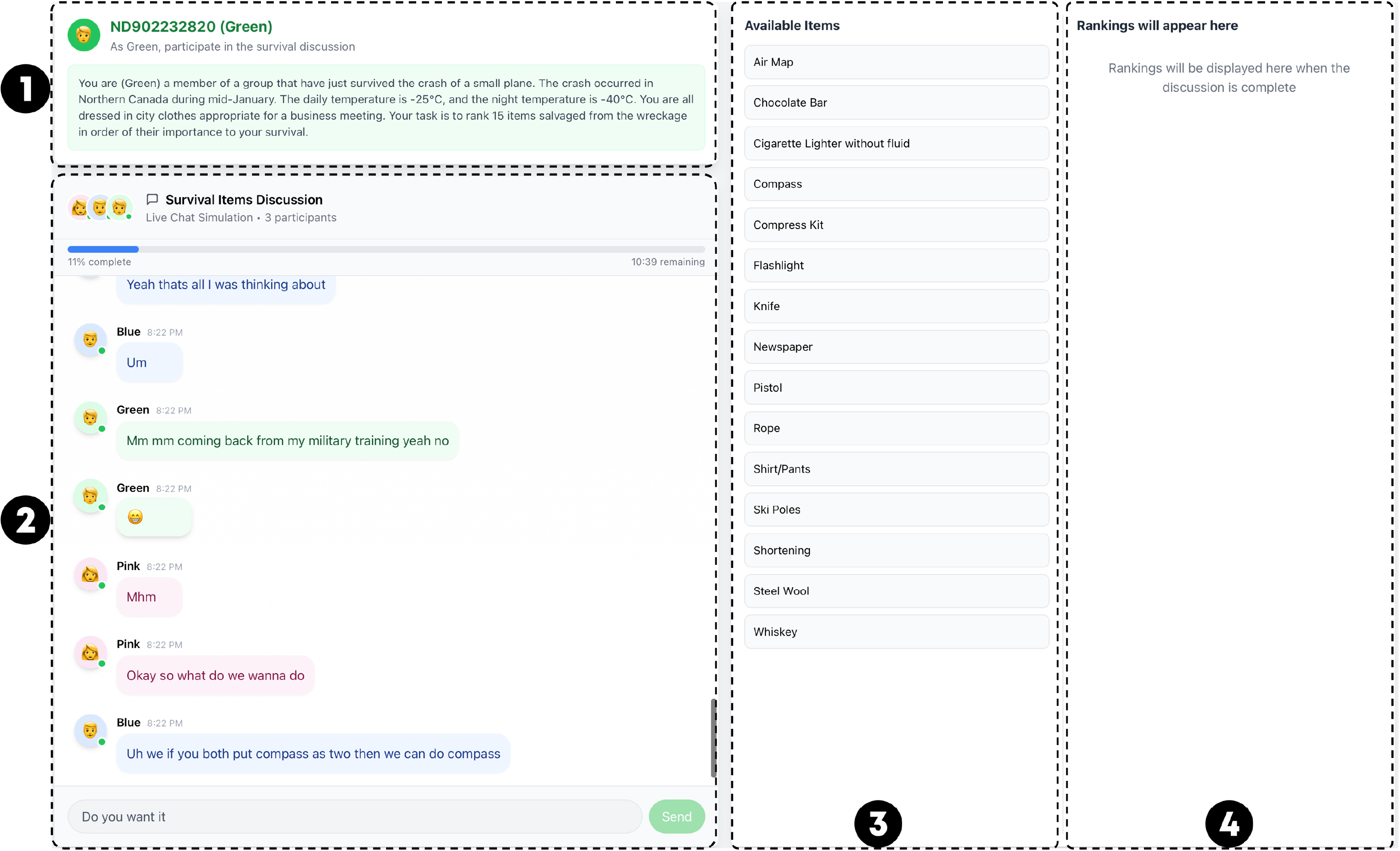}
  \caption{Screenshot of the platform used during our formative study. The participant took the perspective of the “Green” team member, observing communication and interactions among the team as they discussed survival items. Section (1) provides the simulated task description, offering contextual details and instructions. Section (2) displays team members’ discussions and decision-making processes in a simulated real-time chat window. Section (3) presents the list of available survival items that the team will rank collaboratively. After the allocated time ends, the expert rankings are revealed in Section (4) with a performance score.}
  \label{fig:formative_platform_screenshot}
\end{figure*}

\begin{figure*}[!htb]

  \centering
  \includegraphics[width=\textwidth,height=\textheight,keepaspectratio]{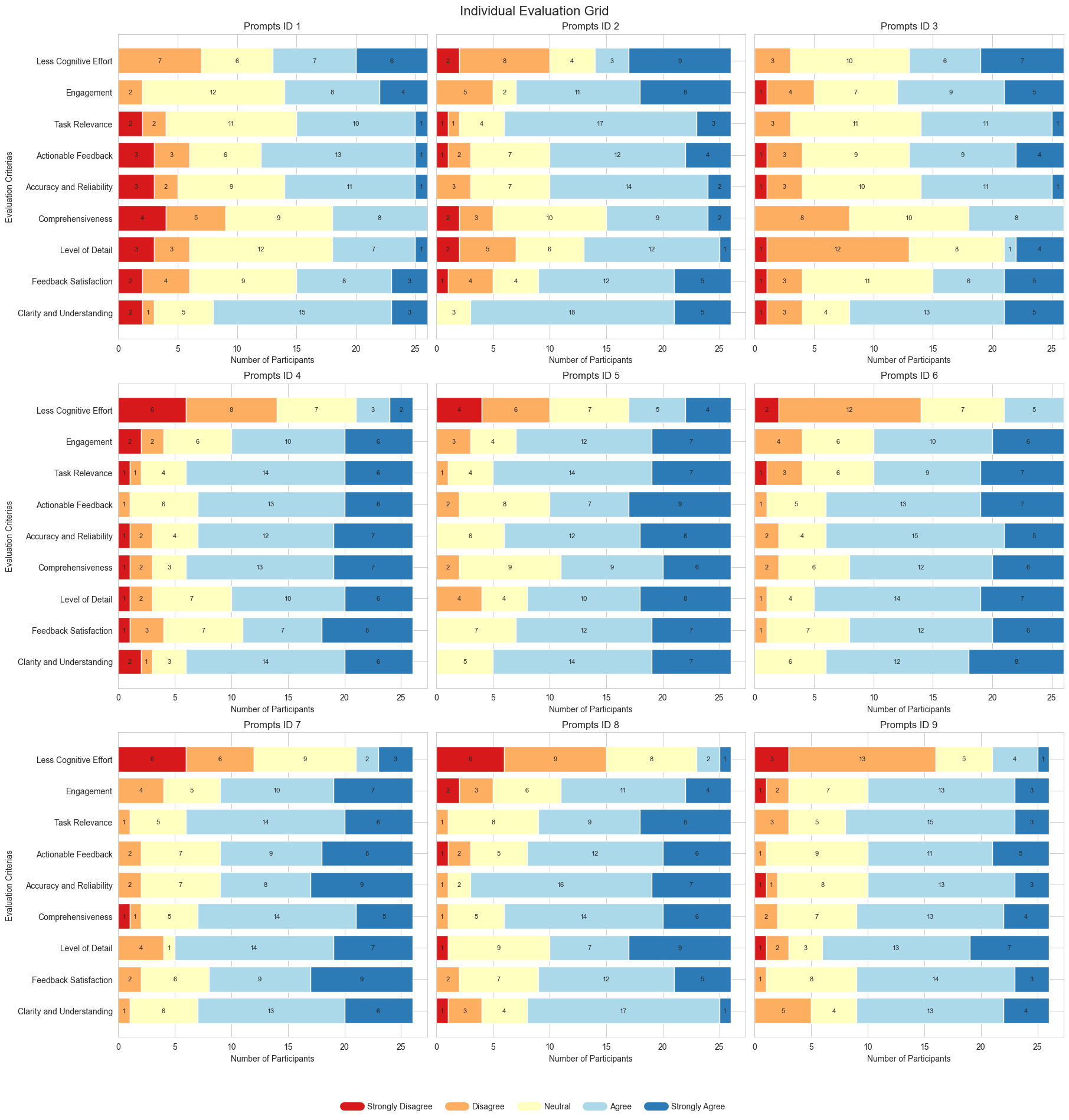}
  \caption{Result from the formative study survey comparing participant rating of different individual feedback messages}
  \label{fig:IndividualEvaluationGrid}
\end{figure*}
\begin{figure*}[!htb]

  \centering
  \includegraphics[width=\textwidth,height=\textheight,keepaspectratio]{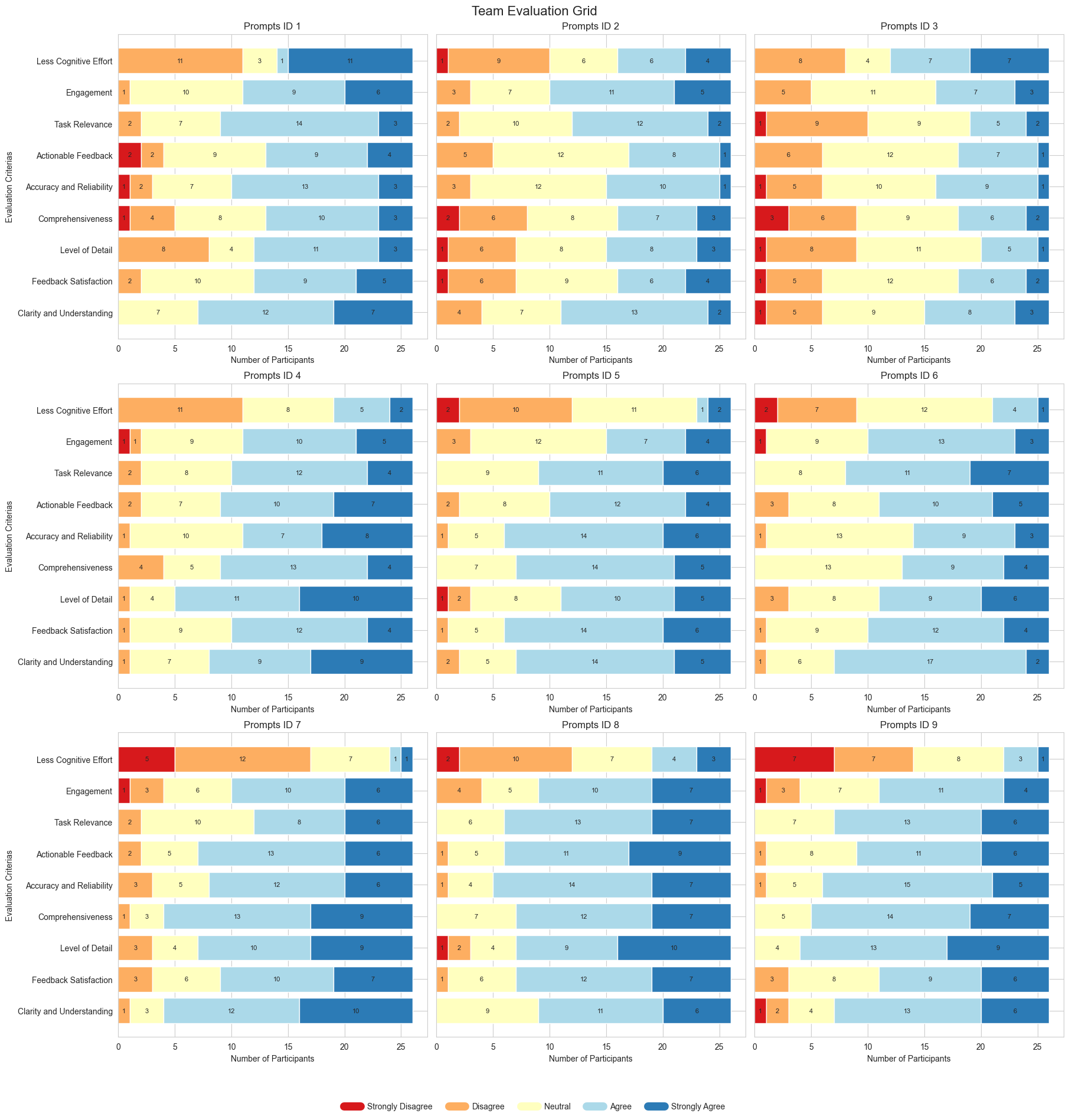}
  \caption{Result from the formative study survey comparing participant rating of different team feedback messages}
  \label{fig:TeamEvaluationGrid}
\end{figure*}

%% file: Appendix/LSM.tex
\section{Language Style Matching (LSM)}
\label{LSM}

LSM is a method that evaluates team members' similarity in using function words during interactions \cite{aafjes2020language}. Studies have shown that LSM reflects how conversational partners align their language styles to achieve a common goal \cite{aafjes2020reciprocal}. For example, Ireland et al. \cite{ireland2010language} showed that nine function words, "articles, auxiliary verbs, conjunctions, high-frequency adverbs, imperatives, personal pronouns, and negations," are unconsciously used during conversations.  In the study, four essay questions were assigned to students, who were required to answer one on a timed website platform. The findings demonstrated that individuals produce utterances that match function words in sentences they have recently heard or read. The results showed that students unintentionally matched the language style of the essay question, and women were more likely to do so.

Building on the framework of Ireland et al. \cite{ireland2010language}, we developed an LSM computation component integrated into \toolname{} to enhance our AI-generated feedback messages presented algorithm \ref{alg:lsm}. We calculate LSM scores based on the team members' communication, providing a statistically grounded foundation for feedback generation. The algorithm uses the \texttt{en\_core\_web\_sm} pipeline provided by SpaCy for tokenization \footnote{\url{https://spacy.io/models/en}}, and the final score is calculated using the following equation:

\[
LSM \textunderscore Score = 1 - \frac{|P_{U} - P_{T}|}{P_{U} + P_{T} + \epsilon}
\\
\]
Where \( P_{U} \) is the proportion of function words in the individual user’s text and \( P_{T} \) is the proportion of function words in the team’s text. \( \epsilon \) is a small constant added to prevent division by zero. The higher LSM scores indicate stronger linguistic engagement between the team members. The theoretical foundation of LSM shows that linguistic style reflects psychological alignment between individuals \cite{niederhoffer2002linguistic, burgoon1995interpersonal}.

\begin{algorithm}

\caption{Language Style Matching (LSM)}\label{alg:lsm}
\KwData{\texttt{Individual\_text}, \texttt{Team\_text}}
\KwResult{\texttt{individual\_LSM\_score}}

\tcp*[l]{Define categories for function words}
\texttt{articles} $\gets$ \{``a'', ``an'', ``the''\}\;
\texttt{personal\_pronouns} $\gets$ \{``i'', ``you'',  \ldots\}\;
\texttt{impersonal\_pronouns} $\gets$ \{``it'', ``its'',  \ldots\}\;
\texttt{prepositions} $\gets$ \{e.g., ``in'', ``to'',  \ldots\}\;
\texttt{auxiliary\_verbs} $\gets$ \{``is'', ``has'', \ldots\}\;
\texttt{adverbs} $\gets$ \{e.g., ``quickly'', ``often'', \ldots\}\;
\texttt{conjunctions} $\gets$ \{``and'', ``or'', \ldots\}\;
\texttt{negations} $\gets$ \{``no'', ``not'',  \ldots\}\;
\texttt{quantifiers} $\gets$ \{``all'', ``any'',  \ldots\}\;

\tcp*[l]{Parse text}
\SetKwFunction{FParsingText}{Parsing\_Text}
\SetKwProg{Fn}{Function}{:}{}
\Fn{\FParsingText{\texttt{text}}}{
  Tokenize \texttt{text} using an NLP model;
  Initialize \texttt{function\_words} as an empty list\;
  \ForEach{token in \texttt{text}}{
    \If{token matches a category condition}{
      Add token to \texttt{function\_words}\;
      Break
    }
  }
  \KwRet \texttt{function\_words}\;
}

\tcp *[l]{Calculate LSM for between two texts (User text and Team text)}
\SetKwFunction{FCalculateLSM}{Calculate\_LSM}
\Fn{\FCalculateLSM{\texttt{User\_text}, \texttt{Team\_text}}}{
  Initialize \texttt{scores} as an empty list\;
  \tcp*[l]{Tokenize and extract function words from both texts}
  \texttt{User\_words} $\gets$ \FParsingText{\texttt{User\_text}}\;
  \texttt{Team\_words} $\gets$ \FParsingText{\texttt{Team\_text}}\;

  \ForEach{category in \texttt{categories}}{
    \tcp*[l]{Compute proportions of function words for both texts}

    Compute $P_{U} \gets \frac{\texttt{count of function words in User\_text}}{\texttt{total words in User\_text}}$ \\
    Compute $P_{T} \gets \frac{\texttt{count of function words in Team\_text}}{\texttt{total words in Team\_text}}$ 

    \texttt{Category\_LSM} $\gets 1 - \frac{|P_{U} - P_{T}|}{P_{U} + P_{T} + \epsilon}$\;
    Append \texttt{Category\_LSM} to \texttt{scores}\;
  }
  \tcp*[l]{Return the overall LSM score as the average across all categories}

  \KwRet \texttt{LSM\_score} $\gets$ Average(\texttt{scores})\;
}

\end{algorithm}


%% file: Appendix/Task.tex
\section{Experimental Tasks}
\label{Experimental_Tasks}

In this section, we describe the three tasks employed in our main user study (see section \ref{User Study}) to assess team engagement, collaboration, and decision-making. Each task introduces a different survival scenario, with the objective of ranking survival items according to their importance in that scenario. The ground truth comes from an expert ranking dependent on the scenario, but team members are not expected to know this ranking. Each team must agree on their final ranking before they can submit their list.

\subsection{Task 1: Survival in Northern Canada \cite{survivalGamePlan}}

\subsubsection{Scenario Details}
You and your team have just survived the crash of a small plane. Both the pilot and co-pilot were killed in the crash. It is mid-January, and you are in Northern Canada. The daily temperature is 25 below zero, and the nighttime temperature is 40 below zero. There is snow on the ground, and the countryside is wooded, with several creeks criss-crossing the area. The nearest town is 20 miles away. You are all dressed in city clothes appropriate for a business meeting.

\subsubsection{Salvaged Items}
\begin{itemize}
    \item \textbf{Ball of steel wool:} Can be used to start a fire for warmth and cooking.
    \item \textbf{Small ax:} Useful for cutting wood to build a shelter or fuel a fire.
    \item \textbf{Loaded .45-caliber pistol:} Can be used for self-defense or hunting small game.
    \item \textbf{Can of Crisco shortening:} Useful as a source of calories or as a fire starter.
    \item \textbf{Newspapers (one per person):} Can be used as insulation or kindling for a fire.
    \item \textbf{Cigarette lighter (without fluid):} Can be used to create sparks for starting a fire.
    \item \textbf{Extra shirt and pants for each survivor:} Useful for layering to prevent hypothermia.
\end{itemize}

\subsubsection{Expert Solution}
\begin{enumerate}
    \item Cigarette lighter (without fluid)
    \item Ball of steel wool
    \item Extra shirt and pants for each survivor
    \item Can of Crisco shortening
    \item Small ax
    \item Newspapers (one per person)
    \item Loaded .45-caliber pistol
\end{enumerate}

\subsection{Task 2: Survival in the Open Ocean \cite{survivalGameSea}}

\subsubsection{Scenario Details}
You and your team have chartered a yacht with an experienced skipper and a two-person crew. A fire breaks out, destroying much of the yacht and its contents. The yacht is slowly sinking, and your location is unclear. The skipper and crew were lost while trying to fight the fire. You are approximately 1000 miles southwest of the nearest landfall.

\subsubsection{Salvaged Items}
\begin{itemize}
    \item \textbf{Case of army rations:} Food supply for sustenance.
    \item \textbf{2 boxes of chocolate bars:} High-energy food source.
    \item \textbf{5 gallon can of water:} Essential for hydration and survival.
    \item \textbf{20 square feet of opaque plastic sheeting:} Useful for collecting rainwater or as a shelter.
    \item \textbf{Fishing kit:} Can be used to catch fish for food.
    \item \textbf{Shaving mirror:} Can be used for signaling for rescue.
    \item \textbf{2 gallon can of oil/petrol mixture:} Potentially useful for starting fires or signaling.
\end{itemize}

\subsubsection{Expert Solution}
\begin{enumerate}
    \item Shaving mirror
    \item 2 gallon can of oil/petrol mixture
    \item 5 gallon can of water
    \item Case of army rations
    \item 20 square feet of opaque plastic sheeting
    \item 2 boxes of chocolate bars
    \item Fishing kit
\end{enumerate}

\subsection{Task 3: Survival on the Moon \cite{survivalGameMoon}}

\subsubsection{Scenario Details}
You are a member of a space crew originally scheduled to rendezvous with a mother ship on the lighted surface of the moon. However, due to mechanical difficulties, your ship was forced to land at a spot some 200 miles from the rendezvous point. During reentry and landing, much of the equipment aboard was damaged, and since survival depends on reaching the mother ship, the most critical items available must be chosen for the 200-mile trip.

\subsubsection{Salvaged Items}
\begin{itemize}
    \item \textbf{Box of matches:} Common matches for starting fires.
    \item \textbf{Food concentrate:} Compact, high-energy food supply.
    \item \textbf{50 feet of nylon rope:} Durable rope for climbing or securing items.
    \item \textbf{Parachute silk:} Can be used as a protective covering or shelter.
    \item \textbf{Two 100 lb. tanks of oxygen:} Essential for breathing in a low-oxygen environment.
    \item \textbf{Stellar map:} Used for navigation and locating the rendezvous point.
    \item \textbf{5 gallons of water:} Crucial for hydration and survival.
\end{itemize}

\subsubsection{Expert Solution}
\begin{enumerate}
    \item Two 100 lb. tanks of oxygen
    \item 5 gallons of water
    \item Stellar map
    \item Food concentrate
    \item 50 feet of nylon rope
    \item Parachute silk
    \item Box of matches
\end{enumerate}

\subsection{Evaluation Formula}
\toolname{} evaluates team performance by calculating a normalized score indicating how closely a team's item ranking aligns with the expert ranking. The evaluation score is computed using the following equation:

\label{Evaluation Formula}
\begin{equation*}
\resizebox{\columnwidth}{!}{$
\text{Score} = \left(\frac{\text{Distance from Worst Ranking} - \text{Distance from Expert Ranking}}{\text{Distance from Worst Ranking}}\right) \times 100
$}
\label{eq:score}
\end{equation*}

where \textit{Distance from Expert Ranking} is computed as the sum of absolute positional differences between the team’s item ranking and the expert ranking, the \textit{Distance from Worst Ranking} is the sum of absolute positional differences between the team’s item ranking and the inverse expert ranking.

\subsection{Human vs. AI feedback comparison}
\label{human-ai-comparison-item}
Please use the rating scale to indicate how much you agree or disagree with the following statements.
\begin{itemize}
    \item \textit{``I found the messages provided by the AI Feedback Bot more helpful than the ones that a human manager could provide.''}
    \item \textit{``A human manager could not provide better insights or details than the AI Feedback Bot.''}
    \item \textit{``I would trust the feedback messages provided by the AI Feedback Bot more than the feedback messages provided by a human manager.''}
    \item (R) \textit{``A human manager’s feedback would generally be more trustworthy than the AI Feedback Bot's feedback.''}
    \item (R) \textit{``The AI Feedback Bot's feedback felt impersonal compared to feedback from a human manager.''}
\end{itemize}

%% file: Appendix/UserStudy.tex
\section{\toolname Prompts}
\label{apandxPrompts}

\subsection{\toolname System Prompt:}

Below is the system prompt utilized in \toolname{}:
\promptbox{
    You are a professional team performance coach specializing in workplace communication. 
    Your role is to analyze discussions, offer constructive feedback, and help improve team cohesion and performance.
    Always be respectful and constructive, and focus on actionable insights.
}

\subsection{\toolname Team Feedback Prompt:}
Below is the team feedback prompt utilized in \toolname{}, which incorporates the following key data points:

\begin{itemize}
  \item$\textcolor{DarkGoldenrod}{TeamConversation}$: Represents the team's conversation script.
  \item$\textcolor{DarkGoldenrod}{TeamSentiment}$: The overall sentiment derived from the team’s conversation.
  \item$\textcolor{DarkGoldenrod}{TotalTeamMember}$: Indicates the total number of team members involved.
  \item$\textcolor{DarkGoldenrod}{TeamTask}$: The task assigned to the team during the session.
\end{itemize}

\promptbox{
    Team Feedback Guide:\\
    KEEP FEEDBACK WITHIN A REASONABLE LENGTH 300 WORDS FOR CLARITY AND ENGAGEMENT.
    
    Use the following Markdown formatting guidelines to enhance the readability and impact of your feedback:
    
    \begin{itemize}
    \item  Use $>$ for headings or to highlight key sections instead of double asterisks $(**)$ or triple hashes.
    \item Use SINGLE asterisks between words to emphasize key points and maintain structure.
    \item Use underscores to italicize both sides of the word or phrase. 
    \end{itemize}

    Objective:
    
    Provide highly detailed and actionable feedback for the team, highlighting their key strengths,
    identifying specific areas for improvement, and offering practical recommendations to enhance collaboration, performance, and alignment with team objectives.

     Key Data Point to Utilize:
     \begin{itemize}
      \item Emotion: Using the given data sentiment data point $\textcolor{DarkGoldenrod}{TeamSentiment}$, Assess the emotion of the conversation and its influence on collaboration.
      \item Topic Coherence: Using the given data points of team conversation $\textcolor{DarkGoldenrod}{TeamConversation}$ and team objective $\textcolor{DarkGoldenrod}{TeamTask}$, team conversation remains aligned with the team task and avoids irrelevant discussion that may lower the team’s effectiveness.
      \item Transactive Memory Systems: Using the given data points of team conversation $\textcolor{DarkGoldenrod}{TeamConversation}$, examine how effectively team members utilize each other’s expertise and shared knowledge.
      \item Collective Pronoun Usage: Using the given data points of team conversation $\textcolor{DarkGoldenrod}{TeamConversation}$, assess the extent to which team members use inclusive language to foster a sense of unity.
      \item Communication Flow: Using the given data points of team conversation $\textcolor{DarkGoldenrod}{TeamConversation}$, identify patterns in turn-taking, response delays, and interruptions to evaluate conversation balance utilizing the timestamps in the data point.
      \item Engagement: Using the given data points of team conversation $\textcolor{DarkGoldenrod}{TotalTeamMemeber}$, evaluate how actively all team members contributed to the discussion.
    \end{itemize}
}
\promptbox{

     Feedback Structure:
   
    Detailed Summary of Goals and Contributions:
     \begin{itemize}
    \item Summarize the team’s goals and evaluate how effectively their discussion aligned with achieving those objectives.  
    \item Example: The team’s primary objective was to develop a strategy for improving customer engagement. Most members maintained alignment with the task, as evidenced. However, a few moments of digression into unrelated topics were noted.
    
    \end{itemize}

    Areas of Strengths: Identify 5-6 key strengths demonstrated by the team. For each strength, provide a specific example that highlights how it was exhibited during the discussion:
     Examples Key Strengths: 
    \begin{itemize}
    \item Engagement: All team members actively participated. For instance, Member A shared a detailed proposal on the topic, which sparked productive discussion.
    \item Positive Sentiment: The team maintained an optimistic tone throughout, fostering a supportive atmosphere. Notably, Member B encouraged collaboration by stating, Let’s combine our ideas to create something impactful.
    \item Collaboration: Ideas were frequently built upon, such as when Member C expanded on Member D’s initial suggestion by adding actionable steps.
    \item Focus on Goals: Frequent objective references helped keep the conversation task-oriented. For example, Member E reminded the group of the deadline during a potential digression.
    \item Creativity and Problem-Solving: Members demonstrated innovative thinking, such as suggesting specific solutions, which was well-received and incorporated into the discussion.
    \end{itemize}

    Areas of Improvement: Identify 5-6 areas where the team can improve. Provide specific examples illustrating these challenges and suggest how they might be addressed.
    Examples of Areas for Improvement:
    \begin{itemize}
    \item Participation Balance: While engagement was generally high, Member F contributed significantly less, which may indicate disengagement or a lack of opportunity to participate.
    \item Turn-Taking: Frequent interruptions and overlapping comments, such as during the discussion on the topic, disrupted the flow of conversation and clarity.
    \item Sentiment Variation: Although the tone was generally positive, comments like 'This idea won’t work' from Member G could have been reframed to maintain morale.
    \item Focus and Alignment: The conversation occasionally drifted off-task, particularly during the brainstorming phase,
     when unrelated topics were introduced.
    \item Clarity and Depth: Certain suggestions, such as the one regarding, lacked sufficient detail to be actionable, requiring further elaboration.
    \end{itemize}

}
\promptbox{
    
    Actionable Steps for Improvement: Offer concrete, practical, and measurable strategies to improve team collaboration and address the identified areas for growth.
    Example recommendations of actionable steps:
     \begin{itemize}
     \item Encourage Balanced Participation:  Assign roles or directly invite quieter members, such as Member F, to share their perspectives. For example, ask, 'What do you think about this approach?'
    \item Enhance Turn-Taking: Establish guidelines for speaking turns to minimize interruptions. 
       Consider using a visual cue, such as raising hands, to signal turns during discussions.
    \item Maintain Positive Sentiment:  Encourage members to frame critiques constructively,  such as suggesting alternatives instead of dismissing ideas outright.
    \item Revisit Goals Regularly: Periodically restate the objectives during discussions to prevent digressions and maintain focus.
    \item Expand on Ideas: Request additional details or examples for unclear suggestions. For instance, during the discussion on a specific topic, ask, Can you elaborate on how this would work in practice?
    \end{itemize}

     Example:
    
    Strengths:  
    \begin{itemize}
     \item *Engagement:* Members A and B actively participated, contributing over 50\% of the discussion.  
     \item  *Positive Sentiment:* Encouraging remarks like 'We’re making progress!' boosted morale.  
     \item  *Collaboration:* Member C built on Member D’s idea about a specific solution, creating a comprehensive plan.  
     \item  *Creativity:* Members demonstrated innovative thinking by proposing specific solutions.  
     \item  *Alignment:* Frequent references to the task goals ensured the team stayed on track.
    \end{itemize}

    Areas for Improvement:  
    \begin{itemize}
     \item *Participation Gaps:* Member F’s limited contributions indicate the need for more inclusivity.  
     \item  *Focus Drift:* The conversation strayed into unrelated topics during a specific moment.  
     \item  *Turn-Taking:* Overlapping comments during specific discussion hindered clarity.  
    \item   *Sentiment Variation:* Negative remarks from Member G dampened collaboration briefly. 
    \end{itemize}

    Actionable Steps for Improvement:
     \begin{itemize}
    \item Maintain a balance between positive reinforcement and constructive criticism.
    \item Ensure feedback is structured, actionable, and aligned with the team’s specific dynamics and performance.
    \item Leverage the provided data to craft a highly detailed, impactful feedback response tailored to the team’s performance and collaboration style.
    \end{itemize}
}

\subsection{\toolname Individual Feedback Prompt:}

Below is the individual feedback prompt utilized in \toolname{}, which incorporates the following key data points:

\begin{itemize}

\item$\textcolor{DarkGoldenrod}{Speaker}$: The identifier of the individual providing the feedback.
\item$\textcolor{DarkGoldenrod}{TeamSentiment}$: The overall sentiment derived from the team’s conversation.
\item$\textcolor{DarkGoldenrod}{MemberSentiment}$: The sentiment specific to the individual’s contributions.
\item$\textcolor{DarkGoldenrod}{TeamTask}$: The task assigned to the team during the session.
\item$\textcolor{DarkGoldenrod}{TeamMemberScript}$: The transcript of the individual’s interactions within the team context.
\item$\textcolor{DarkGoldenrod}{TeamConversation}$: Represents the team's conversation script.
\item$\textcolor{DarkGoldenrod}{TeamMemberName}$: The name of the team member.
\item$\textcolor{DarkGoldenrod}{MemberWordsSpkenPercentage}$: The percentage of words spoken by the individual relative to the entire team.
\item$\textcolor{DarkGoldenrod}{LanguageStyleMatching}$: The degree of similarity between the individual’s language style and that of the team. 
\end{itemize}


\promptbox{
Feedback Guide for $\textcolor{DarkGoldenrod}{Speaker}$: \\
- KEEP FEEDBACK WITHIN A REASONABLE LENGTH 200 WORDS FOR CLARITY AND ENGAGEMENT.
    
    Use the following Markdown formatting guidelines to enhance the readability and impact of your feedback:
    \begin{itemize}
    \item  Use $>$ for headings or to highlight key sections instead of double asterisks $(**)$ or triple hashes.
    \item Use SINGLE asterisks between words to emphasize key points and maintain structure.
    \item Use underscores to italicize both sides of the word or phrase. 
    \end{itemize}

    Objective:
    Provide high-quality, detailed feedback for {speaker} by highlighting their strengths, identifying specific areas for improvement,
    and offering actionable steps based on the provided data.

     Key Data Point to Utilize:
     \begin{itemize}
      \item Emotion: Using the given data sentiment data point of overall team sentiment $\textcolor{brown}{TeamSentiment}$ and the team member sentiment $\textcolor{DarkGoldenrod}{MemberSentiment}$ Assess the emotion of the conversation and its influence on collaboration.
      \item Topic Coherence: Using the given data points of team conversation $\textcolor{DarkGoldenrod}{TeamConversation}$ and the team member script $\textcolor{DarkGoldenrod}{TeamMemberScript}$ and team objective $\textcolor{DarkGoldenrod}{TeamTask}$, evaluate how team conversation remains aligned with the team task and avoids irrelevant discussion that may lower the team’s effectiveness.

       \item Transactive Memory Systems: Using the given data points of team conversation $\textcolor{DarkGoldenrod}{TeamConversation}$ and team member script 
       $\textcolor{DarkGoldenrod}{TeamMemberScript}$, examine how effectively the team member utilizes other’s expertise and shared knowledge.
      \end{itemize}
}
\promptbox{
    \begin{itemize}
      
      \item Collective Pronoun Usage: Using the given data points of team member conversation $\textcolor{DarkGoldenrod}{TeamMemberScript}$, assess the extent to which team members use inclusive language to foster a sense of unity compared to team member conversation data point $\textcolor{DarkGoldenrod}{TeamConversation}$.
      \item Communication Flow: Using the given data points of team conversation $\textcolor{DarkGoldenrod}{TeamConversation}$ and team member name $\textcolor{DarkGoldenrod}{TeamMemberName}$ identify patterns in turn-taking, response delays, and interruptions to evaluate Communication Flow utilizing the timestamps in the data point.
      \item  Team Member engagement: Using this data point $\textcolor{DarkGoldenrod}{MemberWordsSpokenPercentage}$, Evaluate participation and engagement levels.
      \item  Language Style Matching: Using this data point $\textcolor{DarkGoldenrod}{LanguageStyleMatching}$, that provides an indicator of the frequency of function words used by the team member comparing the word distributions, a higher indicates greater linguistic alignment.    
    \end{itemize}

    General Instructions:
    \begin{itemize}
    \item Summarize the speaker's contribution to the team task in terms of effectiveness and alignment with team goals.
     \item Provide specific details on strengths and areas for improvement to make feedback more personalized and actionable.
     \item Clearly outline the challenges and frame them as opportunities for development, offering practical recommendations.
     \item Maintain a neutral, encouraging, and professional tone throughout.
    Use precise and concise language while ensuring sufficient detail for meaningful insights.
    \end{itemize}

    Feedback Structure:
    Summarize $\textcolor{DarkGoldenrod}{TeamMemberName}$'s role in the discussion and how their input aligned with the team's task and goals.
    
    Key Strengths:
    Highlight 4-5 key strengths demonstrated by $\textcolor{DarkGoldenrod}{TeamMemberName}$ during the discussion, Examples:
    \begin{itemize}
    \item  *Sentiment:* Positive attitude, motivation, or enthusiasm.
    \item  *Engagement:* High participation with a total words spoken.
    \item  *Alignment:* Contributions directly advanced the team’s objectives.
    \item  *Language Style Matching:* Communication style harmonized well with the team dynamic.
    \item  Provide specific examples where the speaker excelled, referencing key moments in the conversation. 
    \end{itemize}

    Areas for Improvement:
    Identify 4-5 areas for improvement. Examples:
    \begin{itemize}
    \item *Sentiment:* Address any instances of negative or neutral tones.
    \item *Depth of Contribution:* Expand on points with more details or examples.
    \item *Language Style Matching:* Improve alignment with the team's communication style where applicable.
    - Highlight specific moments where improvement could enhance the speaker’s impact.
    \end{itemize}
}
\promptbox{
    Example 1:
    Summary of Contribution:
    You actively participated in the discussion, contributing significantly to the team’s task.
    Your engagement and enthusiasm were evident throughout the conversation.
    Strengths:
    \begin{itemize}
    \item *Engagement:* You demonstrated active participation with a \%70 of thewords spoken.
    \item *Sentiment:* Your positive tone helped create an encouraging and collaborative environment.
    \item *Task Alignment:* Your comments consistently aligned with the team's objectives, keeping the discussion on track.
    \item *Collaboration:* You acknowledged others' ideas and built upon them effectively, fostering teamwork.
    \end{itemize}

    Areas for Improvement:
    \begin{itemize} 
    \item *Language Style Matching:* At times, your language style differed slightly from the team’s tone, which could lead to minor misunderstandings.
    \item *Depth of Contribution:* Some of your points lacked supporting details, which could make them more impactful.
    \item *Inclusivity:* Encouraging quieter members to share their ideas would further strengthen team dynamics.
    \end{itemize}
    
    Actionable Steps:
    \begin{itemize} 
    \item Aim to align your communication style more closely with the team’s overall tone.
    \item Provide specific examples or reasoning to add depth to your contributions.
    \item Proactively invite quieter team members to contribute by asking for their input.
    \end{itemize}

    Example 2:
    Summary of contribution: Your contributions were insightful and advanced the team’s understanding of team tasks. You showed enthusiasm and worked effectively within the group dynamic.
    
    Strengths:
    \begin{itemize} 
    \item *Sentiment:* Your positive tone motivated the team and boosted morale.
    \item *Engagement:* With a 50\% percentage of words spoken, you actively engaged and contributed valuable ideas.
    \item * Language Style matching: * Your communication style aligned well with the team, fostering a harmonious discussion.
    \item *Problem-Solving:* You proposed practical solutions that were well received by the group.
    \end{itemize}
    
    Areas for Improvement:
    \begin{itemize} 
    \item *Participation Balance:* At times, your input overshadowed quieter team members.
    \item *Emotion:* There were a few instances in which a more neutral tone would have been beneficial.
    \item * Clarity of ideas: * Adding more structure to your responses could enhance clarity and impact.
    \end{itemize}
    
    Actionable Steps:
    \begin{itemize} 
    \item Encourage quieter members to share their perspectives to ensure balanced participation.
    \item Reflect on tone during critical moments to maintain consistency.
    \item Use structured responses to clearly present your ideas.
    \end{itemize}
}

\subsection{\toolname Ranking Evaluation Prompt:}
Below is the ranking evaluation prompt utilized in \toolname{}, which incorporates the following key data points:
\begin{itemize}
\item$\textcolor{DarkGoldenrod}{TeamTask}$: The task assigned to the team during the session.
  \item$\textcolor{DarkGoldenrod}{Ranking}$: The ranking order submitted by the team.
  \item$\textcolor{DarkGoldenrod}{Score}$: The percentage score reflects the team's alignment with the expert solution.
  \item$\textcolor{DarkGoldenrod}{ExpertRanking}$: The ranking order defined by experts.
\end{itemize}

\promptbox{
You are an AI evaluator analyzing a team's performance in a survival simulation task.

\subsection*{Task Context}
Use the following Markdown formatting guidelines to enhance the readability and impact of your feedback:
\begin{itemize}
    \item Use $>$ for headings or to highlight key sections instead of double asterisks $(**)$ or triple hashes.
    \item Use \textit{single asterisks} between words to emphasize key points and maintain structure.
    \item Use \textit{underscores} on both sides of the word or phrase to italicize.
\end{itemize}

\subsection*{Instructions}
Provide an evaluation explanation for the task:

\textbf{Team Task:} $\textcolor{DarkGoldenrod}{TeamTask} $\\
\textbf{Team Submission:} The team has submitted the following ranking: $\textcolor{DarkGoldenrod}{Ranking}$ and received a score of $\textcolor{DarkGoldenrod}{Score}\%$. \\
\textbf{Expert Solution:} The Expert Solution Ranking for the task is: $\textcolor{DarkGoldenrod}{ExpertRanking}$.

Your answer should follow these guidelines:
\begin{itemize}
    \item Do not provide feedback based on their score.
    \item Do not provide feedback on areas unrelated to the ranking choices.
    \item Keep the response concise and focused on the ranking decisions.
\end{itemize}

\subsection*{Examples}
\begin{itemize}
    \item Example 1: The team ranked item \textit{A} in the 1st place, whereas the expert ranked it in the 3rd place.
    \item Example 2: The team ranked item \textit{B} in the 1st place, which was consistent with the expert ranking.
\end{itemize}
}